\documentclass[a4paper,12pt]{article}
\usepackage{amsmath,amssymb}
\usepackage{jheppub}
\usepackage{hyperref}
\usepackage{braket}
\usepackage{graphicx}
\usepackage{mathrsfs}
\usepackage{float}
\usepackage{ulem}
\usepackage{color}

\newcommand{\be}{\begin{equation}}
\newcommand{\ee}{\end{equation}}
\newcommand{\beq}{\begin{equation}}
\newcommand{\eeq}{\end{equation}}

\newcommand{\p}{\partial}

\newcommand{\bea}{\begin{equation}\begin{aligned}}
\newcommand{\eea}{\end{aligned}\end{equation}}
\newcommand{\ba}{\begin{align}}
\newcommand{\ea}{\end{align}}

\newcommand\rref[1]{(\ref{#1})}

\title{Gravitational path integral from the $T^2$ deformation}

                                           \author[a]{Alexandre Belin,}
                                           \author[b]{Aitor Lewkowycz}
                                           \author[a]{and G\'abor S\'arosi}
                                           \affiliation[a]{CERN, Theory Division, 1 Esplanade des Particules \\ Geneva 23, CH-1211, Switzerland}
                                           \affiliation[b]{{Stanford Institute for Theoretical Physics, Department of Physics,\\
Stanford University, Stanford, CA 94305, U.S.A.}}
                                           
                                           \emailAdd{a.belin@cern.ch}
                                           \emailAdd{lewkow@stanford.edu}
                                           \emailAdd{gabor.sarosi@cern.ch}
                                           
\abstract{We study a $T^2$ deformation of large $N$ conformal field theories, a higher dimensional generalization of the $T\bar T$ deformation. The deformed partition function satisfies a flow equation of the diffusion type. We solve this equation by finding its diffusion kernel, which is given by the Euclidean gravitational path integral in $d+1$ dimensions between two boundaries with Dirichlet boundary conditions for the metric. This is natural given the connection between the flow equation and the Wheeler-DeWitt equation, on which we offer a new perspective by giving a gauge-invariant relation between the deformed partition function and the radial WDW wave function.
An interesting output of the flow equation is the gravitational path integral measure which is consistent with a constrained phase space quantization.
Finally, we comment on the relation between the radial wave function and the Hartle-Hawking wave functions dual to states in the CFT, and propose a way of obtaining the volume of the maximal slice from the $T^2$ deformation.
}

\begin{document}
\begin{flushright}
\hfill{\tt CERN-TH-2020-085}
\end{flushright}

\maketitle

\section{Introduction}

The AdS/CFT correspondence \cite{Maldacena:1997re} relates theories of quantum gravity in asymptotically Anti-de Sitter space to conformal field theories in one less dimension. The conformal field theory is often thought of as living on the conformal boundary of the asymptically AdS spaces where asymptotic boundary conditions are imposed. Following the standard extrapolate dictionary \cite{Gubser:1998bc,Witten:1998qj}, CFT sources and expectation values of local operators are related to asymptotic values of bulk fields. The AdS setup nicely avoids the ordeal of trying to define local observables in quantum gravity, since the observables (in this case the set of CFT correlation functions) all live in a region where gravity is switched off and where local operators are perfectly well-defined: the asymptotic boundary.

While this has been exploited to our great advantage in AdS/CFT, it is also a burden: One would also like to understand how to define observables at some finite distance in the bulk, which should make sense at least at the semi-classical level (for example, one can formulate an approximate definition of local operators in the bulk by smearing CFT operators \cite{Hamilton:2005ju}). More generally, one can ask how to define quantum gravity with boundary conditions at some finite distance. The radial direction in AdS is related to a UV-cutoff in the CFT providing an RG-perspective on bulk slices at different radial positions \cite{deBoer:1999tgo}, which enabled early attempts at formulating a finite-cutoff holography \cite{Heemskerk:2010hk,Faulkner:2010jy}. An important insight of \cite{Heemskerk:2010hk} was that the Wilsonian action in holographic RG necessarily contains double trace couplings. These couplings are generated as one splits the path integral in the bulk along some radial surface, because there is an extra integral for fields on the surface compared to a conventional Wilsonian splitting in momentum space.

Recently, some further progress has been made on understanding holography and quantum gravity with Dirichlet boundary conditions in the context of AdS$_3$/CFT$_2$ \cite{McGough:2016lol,Kraus:2018xrn,Guica:2019nzm}. The set-up is to consider an integrable irrelevant deformation of two-dimensional CFTs known as a $T\bar{T}$ deformation \cite{Zamolodchikov:2004ce,Smirnov:2016lqw,Cavaglia:2016oda} (see for example \cite{Jiang:2019hxb} for a review), and apply it to holographic CFTs. Similar ideas have been pursued in other dimensions, for example for AdS$_2$ in  \cite{Gross:2019ach,Gross:2019uxi,Iliesiu:2020zld,Stanford:2020qhm} or for $d>2$ in \cite{Bonelli:2018kik,Taylor:2018xcy,Hartman:2018tkw}. For $d>2$, it is still an open question how to define the deformation outside the semi-classical large $N$ limit and we will not attempt to answer this question here. 

In this paper, we study the RG flow equation under a $T^2$ deformation introduced in \cite{Hartman:2018tkw}. Because of its double-trace nature, the flow equation is of diffusion type (see \cite{Cardy:2018sdv,Datta:2018thy} in the 2d case), namely it contains a first derivative with respect to the Wilsonian cutoff, but second derivatives with respect to the sources. It can therefore be solved by a diffusion kernel, which means the flow equation has a solution of the form
\beq
\label{eq:generalkernel}
Z^{(\lambda)}[\gamma]=\int \mathcal{D}\gamma' L[\lambda,\gamma,\lambda',\gamma']Z^{(\lambda')}[\gamma'],
\eeq
where the $Z$ are generating functions for the deformed theories, $\lambda$ is the flow parameter and $\gamma$, $\gamma'$ are background metrics for the field theory. The solution depends on one piece of initial data, which in the case of $\lambda'\rightarrow 0$, is a CFT generating function. In two dimensions, the kernel is a very simple function which is known due to the work of Freidel \cite{Freidel:2008sh} (see \cite{McGough:2016lol,Mazenc:2019cfg} in the context of the $T\bar T$ flow). 

The diffusion type smearing with a kernel as in  \eqref{eq:generalkernel} highlights the peculiar nature of the cutoff in holographic RG. Instead of directly suppressing high frequency modes, the cutoff is provided by smearing the background metric on a scale determined by $\lambda$. This makes it impossible to resolve distances smaller than this scale in the deformed theory,\footnote{This is related to the idea of regulating by a gradient flow \cite{Luscher:2010iy,Shyam:2018sro}.} and explains the difficulty in defining the cherished observables of local quantum field theory such as local correlation functions or entanglement entropy in the $T^2$ deformed theory. Nevertheless, some progress has been achieved in this direction \cite{Cardy:2019qao,Donnelly:2018bef,Banerjee:2019ewu,Murdia:2019fax}.

In higher dimensions, the kernel \eqref{eq:generalkernel} only has a simple form when $\lambda \approx \lambda'$, in which case it is just a Gaussian, giving rise to a Hubbard-Stratonovich representation of the $T^2$ deformation \cite{Hartman:2018tkw}. We will proceed to write a formula for the kernel corresponding to finite deformations by iterating the infinitesimal kernel. The number of iterations corresponds to the radial depth in the bulk. Via a sequence of change of variables, we will show that the iterated kernel is given by a path integral in Euclidean Einstein gravity in $d+1$ dimensions between two surfaces with Dirichlet boundary conditions on the induced metric:
\beq
\label{eq:pathintkernel}
L[\lambda,\gamma,\lambda',\gamma'] = \int_{\substack{q_{IR}=\frac{\gamma}{\lambda^{2/d}} \\ q_{UV}=\frac{\gamma'}{{\lambda'}^{2/d}}}} \mathcal{D}m(g) e^{\frac{1}{16\pi G_N}[\int_{\mathcal{M}} \sqrt{g}(R-2\Lambda)+2\int_{\partial \mathcal{M}}  \sqrt{q}K-\frac{2(d-1)}{\ell}\int_{\partial \mathcal{M}} \sqrt{q}]},
\eeq
where $\mathcal{M}$ is a $d+1$ manifold with two boundaries, $q$ is the induced metric on the boundaries, with $q_{UV}$ and $q_{IR}$ corresponding to the two separate surfaces. We will comment on the measure $\mathcal{D}m(g)$ below.
The formula \eqref{eq:pathintkernel} has a very simple dependence on $\lambda, \lambda'$, which appear only in the dictionary between the induced metric and the QFT background metric.\footnote{We have suppressed some holographic counter terms in \eqref{eq:pathintkernel} for simplicity. In odd boundary dimensions, these also depend only on $q$ but in even dimensions they introduce extra dependence on $\lambda$, related to the conformal anomaly. We will discuss this in the main text.} In particular, $\lambda$ can be position dependent. Note that the action in \eqref{eq:pathintkernel} comes with the ``wrong" sign, a fact anticipated in \cite{Heemskerk:2010hk}. This is necessary in order to interpret the left hand side in \eqref{eq:generalkernel} as a bulk path integral with a finite Dirichlet boundary, since in this case, the role of the kernel is to ``remove" part of the bulk path integral. 

Of course, the appearance of the gravitational path integral is not surprising given the connection of the $T^2$ flow equation to the Wheeler-DeWitt (WDW) equation \cite{McGough:2016lol,Hartman:2018tkw}, which follows from combining the flow equation with a scaling Ward identity. We will revisit this connection from a new perspective by giving the precise (gauge independent) relation between the WDW wave function and the deformed partition function. This will trivialize the role of the conformal anomaly, and explain how initial data of the first order flow equation and the second order Wheeler-DeWitt equation are related. 

Let us return to the resulting measure $\mathcal{D}m(g)$. In general, the question of the measure for the gravitational path integral is a subtle problem. Even though this path integral can only be regarded as defining an effective field theory, the effects of the measure show up in loops that we are supposed to trust bellow the scale of new physics. A snapshot of different proposals for the measure is \cite{Misner:1957wq,Leutwyler:1964wn,Faddeev:1973zb,Fradkin:1974df,Fradkin:1977hw,Polyakov:1981rd,Mazur:1989by}. In solving the flow induced by the $T^2$ deformation, we find a measure $\mathcal{D}m(g)$ that can be interpreted as coming from integrating out the momentum from a Hamiltonian path integral with flat measure $\mathcal{D}q_{ij} \mathcal{D}P^{ij} \mathcal{D}N \mathcal{D}N^i$, where $q_{ij}$ and $P^{ij}$ are the conjugate variables in the ADM formulation \cite{Arnowitt:1959ah}, while $N$ and $N^i$ are the lapse and shift (we refer to \cite{Faddeev:1973zb,Fradkin:1974df,Han:2009bb} for discussions of the Hamiltonian path integral in GR). As explained in \cite{Han:2009bb}, this measure is anomalous under radial diffeomorphisms but invariant under gauge transformations generated by the constraints via the Poisson bracket. These two only agree on-shell. In particular, $\mathcal{D}m(g)$ is different than the diffeomorphism invariant measure \cite{Polyakov:1981rd,Mazur:1989by}, used e.g. on the string worldsheet. 

While the gravitational path integral in \eqref{eq:pathintkernel} is appealing, its connection to AdS/CFT is not completely clear. In particular, the form of the kernel just follows from the form of the flow equation which is well defined whenever the $T^2$ operator can be defined, which only requires large $N$ and not strong coupling. Even at strong coupling, most holographic theories (e.g. $\mathcal{N}=4$ SYM) have light fields  interacting with the metric, dual to single trace operators. In both of these cases, the bulk is not just Einstein gravity, nevertheless we can still turn on the $T^2$ deformation and generate the path integral in \eqref{eq:pathintkernel}. This suggests that the bulk generated in \eqref{eq:pathintkernel} is ``fake", or at least that there are extra conditions on the flow that must be satisfied in order for it to really describe the bulk. We will speculate on possible answers to this question.

In this work, we restricted our attention to the $T^2$ deformation while sourcing only the background metric. Following \cite{Heemskerk:2010hk,Kraus:2018xrn,Hartman:2018tkw}, we expect that turning on appropriate double-trace deformations for other single-trace operators gives rise to diffusion kernels calculated by path integrals involving gravity coupled to matter. We leave this for future work.

Finally, we will offer some thoughts on a related object, the Hartle-Hawking wave function \cite{Hartle:1983ai} of states in AdS gravity. The Hartle-Hawking wave function is a wave function on the intrinsic geometry of initial data (Cauchy) slices. It is calculated by a Euclidean path integral, and in the case of asymptotically AdS gravity, the Euclidean manifold has a boundary. Therefore, one calculates this wave function by putting boundary conditions on both the initial data surface and the asymptotic Euclidean boundary. The latter correspond to sources in the CFT preparation of the state, so these wave functions are dual to Euclidean path integral states \cite{Skenderis:2008dg,Botta-Cantcheff:2015sav,Marolf:2017kvq,Belin:2018bpg} in the CFT. We will argue that the radial wave function has built in knowledge of these HH wave functions and that the HH wave functions can be thought of as overlaps between Euclidean path integral states and metric ``eigenstates". The metric eigenstates are defined from linear combinations of boundary path integral states using the diffusion kernel \eqref{eq:pathintkernel}. Using this picture, we propose a formula for the volume of the extremal slice from the $T^2$ deformation.

The paper is organized as follows: We discuss the $T^2$ flow equation, its connection to the WDW equation and its solution in terms of a diffusion kernel in sec. \ref{sec:2}. We sketch the connection to the Hartle-Hawking wave functions and the volume of the maximal slice in sec. \ref{sec:3}. We end with a discussion of some open problems in sec. \ref{sec:discussion}.

\section{The $T^2$ flow and the bulk path integral}
\label{sec:2}

In this section, we will analyse the flow equation introduced in \cite{Hartman:2018tkw}. This flow equation describes a purely field theoretic deformation in terms of an effective field theory operator for large $N$ CFTs, even though \cite{Hartman:2018tkw} derives it from the bulk. The strategy in \cite{Hartman:2018tkw} to obtain the deforming operator is to relate its expectation value to the trace of the stress tensor via a scaling Ward identity, and then express this trace using the holographic stress tensor \cite{Balasubramanian:1999re} and the Hamiltonian constraint. Because of this, the flow equation encodes bulk equations of motion for gravity, and contains explicit dependence on background fields (sources) and this dependence is partially encoded by holographic counter terms.

We will start by reciting the flow equation from \cite{Hartman:2018tkw}, with a slight modification in the treatment of counter terms and the conformal anomaly, after which we derive the WDW equation from it. In some sense, this is running the argument of \cite{Hartman:2018tkw} backwards, but the process will highlight that the relation between the deformed partition function and the radial WDW wave function is gauge invariant, even though the deformation was originally obtained in the Fefferman-Graham gauge. We will then analyse this relation via a simple toy example, explaining how initial data for the flow and the WDW equation are related, and giving a new perspective on holographic counter terms. After this, we move on to derive the diffusion kernel for the flow equation. We do this in several steps of increasing complexity.

For most of what follows, we will use the convention $16\pi G_N=1$ except when explicitly mentioned.

\subsection{$T^2$ deformation and the WDW equation}

We wish to analyze flow equations of the form considered in \cite{Hartman:2018tkw}. 
\beq
\label{eq:Xdef}
 \frac{d}{d \lambda} \log Z_{\rm ren}^{(\lambda)}(\gamma) = \int d^d x \sqrt{\gamma} \langle X\rangle ,
\eeq
where the deforming operator is
\beq
\label{eq:Hartmanflow}
X=G^{(\gamma)}_{ijkl}\left( T^{ij}+\frac{2}{\sqrt{\gamma}}\frac{\delta S_{c.t}}{\delta \gamma_{ij}} \right)\left( T^{kl}+\frac{2}{\sqrt{\gamma}}\frac{\delta S_{c.t}}{\delta \gamma_{kl}} \right) + \frac{1}{\sqrt{\gamma}}\left(a_d \frac{1}{\lambda}\sqrt{q}R_q+\frac{\delta S_{c.t}}{\delta \lambda } \right) 
\eeq
 where $G^{(\gamma)}_{ijkl}=\frac{1}{2}(\gamma_{ik}\gamma_{jl}+\gamma_{jk}\gamma_{il})-\frac{1}{d-1}\gamma_{ij}\gamma_{kl}$ and $q_{ij}=\lambda^{-2/d}\gamma_{ij}$ and $R_q$ is the Ricci scalar of $q_{ij}$. We will leave the constant $a_d$ unspecified for now and we will see how it is determined shortly. The action $S_{c.t}$ is the holographic counter term action \cite{Henningson:1998gx,Balasubramanian:1999re,deHaro:2000vlm}, but without the boundary cosmological constant. That is, we only include terms in our definition of $S_{c.t}$ that contain derivatives of the metric. In odd dimensions, it depends only on $q_{ij}=\lambda^{-2/d}\gamma_{ij}$ but in even dimensions it includes an anomaly term \cite{Henningson:1998gx,deHaro:2000vlm}
 \beq
 \label{eq:ct1}
 S_{c.t.}[\gamma,\lambda]=\tilde S_{c.t.}[\lambda^{-2/d}\gamma]+\frac{1}{d} \int \log \lambda\sqrt{\gamma} A_d[\gamma],
 \eeq
 where $A_d[\gamma]$ is the local conformal anomaly.\footnote{Note that ref. \cite{Hartman:2018tkw} has a slightly different flow equation than our equation \eqref{eq:Hartmanflow} in the treatment of counter terms. We will explain this difference at the end of this section.}

This flow equation applies for the usual partition function, that we denote by $Z_{\rm ren}^{(\lambda)}[\gamma]$, standing for ``renormalized". The CFT partition function corresponds to zero deformation, i.e. $Z_{CFT}[\gamma]=Z_{\rm ren}^{(\lambda=0)}[\gamma]$. Let us also introduce a bare partition function
\beq
\label{eq:bare-renorm}
Z^{(\lambda)}[\gamma] = e^{S_{c.t}[\gamma,\lambda]}Z_{\rm ren}^{(\lambda)}[\gamma].
\eeq
This bare partition function is finite for $\lambda \neq 0$, and it is a natural object from an RG perspective when a UV regulator is present, for example if we flow to the continuum with a lattice system. In such cases, the bare partition function usually diverges in the continuum limit and needs counter terms to be well-defined, but it is the natural object to consider in the presence of the regulator (see \cite{Caputa:2019pam} for a similar discussion).

The flow equation for the bare partition function is simpler, since it does not contain the counter term variations anymore\footnote{We note that similar equations have appeared before in the holographic RG context in \cite{Mansfield:1999kk,Heemskerk:2010hk}.}
 \beq
 \label{eq:T2flow1}
 \frac{\delta}{\delta \lambda} Z^{(\lambda)}(\gamma) = \frac{4}{\sqrt{\gamma}}G^{(\gamma)}_{ijkl}\frac{\delta}{\delta \gamma_{ij}}\frac{\delta}{\delta \gamma_{kl}}Z^{(\lambda)}(\gamma)  + a_d \frac{1}{\lambda}\sqrt{q}R_q Z^{(\lambda)}(\gamma) .
 \eeq
In this equation we have now allowed for a spatially varying deformation parameter $\lambda(x)$. This is a diffusion equation with potential, which can be solved formally by a diffusion kernel that we will explicitly determine in this paper.

It is clear that the bare partition function $Z$ should be related to the (radial) WDW wave function \cite{deBoer:1999tgo}, but the latter should only depend on the metric $\gamma$ and not the flow parameter $\lambda$. Introducing $\lambda$ is kind of a useful redundancy, since the diff-invariant quantity that naturally keeps track of the radial position is the scale $\sqrt{\gamma}$, see e.g. \cite{Kraus:2018xrn}. To go to the WDW wave function $\Psi$, we notice that e.q. \eqref{eq:T2flow1} preserves the non-anomalous scaling Ward identity
\beq
\label{eq:scaling}
Z^{(\alpha^{-d}\lambda)}[\alpha^{-2}\gamma]=Z^{(\lambda)}[\gamma].
\eeq
Crucially, \eqref{eq:scaling} is satisfied by conformal initial data to the flow, since near the conformal point $\lambda\approx 0$, the conformal anomaly cancels in the bare partition function \eqref{eq:bare-renorm} with the contribution from the counter term \eqref{eq:ct1}, see appendix \ref{app:anomaly}.
This suggests to make a scaling ansatz for $Z$ along the lines of \cite{Freidel:2008sh}:
\beq
\label{eq:frompartfunctowavefunc}
Z^{(\lambda)}[\gamma]=\Psi[\gamma/\lambda^{\frac{2}{d}}]
\eeq
This fixes the notation for the three main objects of interest in this paper, namely the bare and renormalized partition functions $Z$ and $Z_{\text{ren}}$, as well as the WDW radial wave-function $\Psi$.

Putting this ansatz into the flow equation \eqref{eq:T2flow1}, the $\lambda$ dependence cancels and we get
 \beq
 \Big[\frac{2}{d}q_{ij} \frac{\delta}{\delta q_{ij}}+\frac{4}{\sqrt{q}}G^{(q)}_{ijkl}\frac{\delta}{\delta q_{ij}}\frac{\delta}{\delta q_{kl}}+a_d \sqrt{q}R_q \Big]\Psi[q_{ij}]=0.
 \eeq
 This is the WDW equation in slight disguise. We can put it back into more conventional form following \cite{McGough:2016lol}. We first need to do a rescaling of the metric (so that it matches the induced metric that will later come out from the path integral derivation), $q_{ij} \mapsto \left(  \frac{4d}{\ell} \right)^{2/d} q_{ij}$. The new equation for $\Psi$ as a function of this rescaled $q$ is
  \beq
  \label{eq:WDW0}
 \Big[\frac{2}{\ell}q_{ij} \frac{\delta}{\delta q_{ij}}+\frac{1}{\sqrt{q}}G^{(q)}_{ijkl}\frac{\delta}{\delta q_{ij}}\frac{\delta}{\delta q_{kl}}+ \frac{1}{4}a_d\left( \frac{4d}{\ell} \right)^{2\frac{d-1}{d}} \sqrt{q}R_q \Big]\Psi[q_{ij}]=0.
 \eeq
 Now we want to complete squares in the derivatives. Introducing
 \beq
 \label{eq:momentumoperator}
 \hat \pi^{ij} = \frac{\delta}{\delta q_{ij}}-\frac{d-1}{\ell}\sqrt{q}q^{ij}, 
 \eeq
 we have that 
   \beq
     \label{eq:WDW1}
 \Big[\frac{1}{\sqrt{q}}G^{(q)}_{ijkl}\hat \pi^{ij}\hat \pi^{kl}+ \frac{1}{4}a_d\left( \frac{4d}{\ell} \right)^{2\frac{d-1}{d}} \sqrt{q}R_q +\frac{d(d-1)}{\ell^2}\sqrt{q} + \text{contact terms}\Big]\Psi[q_{ij}]=0,
 \eeq
and we must choose $a_d$ so that $\frac{1}{4}a_d\left( \frac{4d}{\ell} \right)^{2\frac{d-1}{d}}=1$ in order to recover the WDW equation. The part ``contact terms" refers to things proportional to $[\hat \pi(x),q(x)] \sim \delta(0)$. These are ordering ambiguities in the WDW equation from a canonical quantization point of view. One might be tempted to say that the $T^2$ deformation fixes these ambiguities (so that we need to quantize the WDW equation in a way that we get \eqref{eq:WDW0} without contact terms), but since the point splitting regularization of the $T^2$ operator is not completely understood, there could in principal be other contact terms induced by it.
 
  Note that there is nothing special in the shifted identification \eqref{eq:momentumoperator}, it is only there because we equated $Z$ with $\Psi$ without stripping the leading counter term, that is, the boundary cosmological constant. We could equivalently just use the identity
 \beq
 \label{eq:cantraf}
  \frac{\delta}{\delta q_{ij}}=e^{2\frac{d-1}{\ell} \int \sqrt{q} }\left(  \frac{\delta}{\delta q_{ij}}+\frac{d-1}{\ell}\sqrt{q}q^{ij}\right)e^{-2\frac{d-1}{\ell} \int \sqrt{q} },
 \eeq
 set $\hat \pi^{ij}= \frac{\delta}{\delta q_{ij}}$ and regard the flow equation for the rescaled wave function $\Psi \mapsto e^{-2\frac{d-1}{\ell} \int \sqrt{q} }\Psi$. This completes the square the same way, therefore we still have the same contact terms in going from \eqref{eq:WDW0} to \eqref{eq:WDW1}.
 
 Of course the above derivation is just running backwards the main argument of ref. \cite{Hartman:2018tkw}. The key new point we will need later is that the WDW wave function is related to the bare partition function via the scaling relation \eqref{eq:frompartfunctowavefunc}, which is now a gauge invariant statement, since the wave function $\Psi$ obeys the constraints\footnote{The momentum constraint is trivially equivalent with the conservation of the bare stress tensor, or coordinate invariance of \eqref{eq:bare-renorm}.} and only depends on the intrinsic geometry of the surface.
 
 Let us return to comment on how our flow equation \eqref{eq:Hartmanflow} slightly differs from the one written in \cite{Hartman:2018tkw}. In \cite{Hartman:2018tkw}, only the anomaly free counter term action $\tilde S_{c.t}$ appears in \eqref{eq:Hartmanflow} via ${\tilde C}^{ij}=\frac{2}{\sqrt{q}}\frac{\delta \tilde S_{c.t.}}{\delta q_{ij}}$ and $\frac{\delta \tilde S_{c.t}}{\delta \lambda }$ expressed via ${\tilde C}^i_i$, in other words, their flow equation does not include the variations of the $\log \lambda$ terms in \eqref{eq:ct1}. For the $\gamma_{ij}$ variations this is harmless, since the $\gamma_{ij}$ variation of the anomaly action is zero in $d=2$, and scheme dependent in $d=4,6$ as explained in \cite{deHaro:2000vlm}. Here we choose a scheme such that these variations are included in \eqref{eq:Hartmanflow} so that we can treat every dimension in a unified manner and work with a non-anomalous Ward identity for our bare stress tensor. On the other hand, the $\lambda$ variation of the $\log \lambda$ term does make a difference compared to \cite{Hartman:2018tkw}, for example in $d=2$ (with the choice of $a_d$ defined after \eqref{eq:WDW1}) we have $X=G^{(\gamma)}_{ijkl}T^{ij}T^{kl}\equiv T{\bar T}$ because the Ricci term cancels with the counterterm variation in \eqref{eq:Hartmanflow}.
Note that for the holographic Brown-York stress tensor to be finite, these log divergences \textit{must} be subtracted \cite{deHaro:2000vlm}.

 \subsection{First order versus second order equations, and need for counter terms}
 \label{sec:toyflows}
 
We have seen that solutions of the $T^2$ flow equation \eqref{eq:T2flow1} that obey the scaling relation \eqref{eq:scaling} also solve the WDW equation. This suggests that we can generate solutions to the WDW equation by solving a diffusion type equation. A confusing point is that the diffusion equation has one initial data, while a second order equation like the WDW equation has two independent solutions, typically with two different type of growths as we take the scale factor to infinity \cite{Freidel:2008sh}. Let us start by understanding this issue via a toy example, that will also highlight the importance of the counter terms. 

The toy example will be the 1d the diffusion equation
\beq
\partial_t Z(t,x) = \partial_x^2 Z(t,x).
\eeq
The equation preserves the scaling property $Z(\alpha^2 t,\alpha x)=Z(t,x)$, which is analogous to the Ward identity \eqref{eq:scaling}. This suggests that there are special solutions $Z(t,x)=\psi(x/\sqrt{t})$. Of course, not all solutions are of this form. The scaling property puts constraint on the initial data
\beq
Z(0,\alpha x) = Z(0,x) \quad \rightarrow \quad Z(0,x)=c_1+c_2 \Theta(x) \quad \text{or} \quad Z(0,x)=\infty.
\eeq
Indeed, one can plug $Z(t,x)=\psi(x^2/t)$ into the diffusion equation to obtain 
\beq
\frac{a}{2}\psi'(a)+ \psi''(a)=0
\eeq
which is a second order equation, and in our example is analogous to the WDW equation. The general solution is $\psi=c_1+c_2 \text{Erf}(a/2)$, which as a solution of the diffusion equation, indeed corresponds to the initial data $Z(0,x)=c_1+c_2 \Theta(x)$. 

For gravity, we will want to think about $x$ as the metric and $t$ as some ``radial coordinate" that is redundantly introduced to keep track of the scale factor. In that case, the situation where the scaling symmetry $\psi(0,\alpha x) = \psi(0,x)$ enforces $\psi(0,x)=\infty$ will be more relevant: we can only give initial data for the diffusion in an asymptotic sense, i.e. at the conformal boundary. To handle this, it is convenient to introduce a ``counter term formalism". Let us illustrate this via a slight generalization of the toy example. We want to consider diffusion with the scale invariant potential
\beq
\label{eq:toywithpot}
\partial_t Z = \partial_x^2 Z - \frac{1}{t}V(\frac{x^2}{t}) Z.
\eeq
This equation still preserves the scaling property $Z(\alpha^2 t,\alpha x)=Z(t,x)$. On the scale invariant ansatz $Z(t,x)=\psi(x/\sqrt{t})$ we get again a second order equation for $\psi(a)$. Writing $\psi(a)=e^{-a^2/8}g(a)$ to eliminate the first derivative, in analogy to the canonical transformation between the two forms of the WDW equation \eqref{eq:WDW0} and \eqref{eq:WDW1}, the equation becomes
\beq
g''(a)-\frac{1}{16}[4+a^2-16 V(a^2)]g(a)=0.
\eeq
To be very concrete, let us pick a potential that gives an equation relevant for a mini superspace WDW equation \cite{Caputa:2018asc,Caputa:2019pam}, $V(a^2)=1/4-a+a^2/16$. This results in the equation
\beq
a g(a)-g''(a)=0,
\eeq
whose solutions are the Airy functions
\beq
g(a)=C_1 \text{Ai}(a)+C_2 \text{Bi}(a).
\eeq
Since $a=x/\sqrt{t}$, the small $t$ regime means $a\rightarrow \infty$, where the above solution has asymptotic form
\bea
g(a) &=\frac{C_1}{2\sqrt{\pi}} a^{-1/4}e^{-\frac{2}{3}a^{3/2}} + \frac{C_2}{\sqrt{\pi}} a^{-1/4}e^{\frac{2}{3}a^{3/2}}, 
\eea
i.e. there is a decaying and a growing mode. This gives the small $t$ asymptotics of the solution to the corresponding diffusion equation. The key point is that this solution cannot be reproduced from the original diffusion equation \eqref{eq:toywithpot} with regular initial data at $t=0$. We can however obtain each branch of the two solutions from a diffusion equation by introducing ``counter terms". That is, we reintroduce $a^2=x^2/t$ and define
\bea
\label{eq:toybranches}
Z^{\rm IR}_{\rm ren}(t,x) &=(x^2/t)^{\frac{1}{8}}e^{\frac{1}{8}\frac{x^2}{t}-\frac{2 x^{3/2}}{3 t^{3/4}}}Z(t,x)= (x^2/t)^{\frac{1}{8}}e^{-\frac{2 x^{3/2}}{3 t^{3/4}}}\text{Bi}(x/\sqrt{t})\\
Z^{\rm UV}_{\rm ren}(t,x) &=(x^2/t)^{\frac{1}{8}}e^{\frac{1}{8}\frac{x^2}{t}+\frac{2 x^{3/2}}{3 t^{3/4}}}Z(t,x)=(x^2/t)^{\frac{1}{8}}e^{\frac{2 x^{3/2}}{3 t^{3/4}}}\text{Ai}(x/\sqrt{t}).
\eea
These functions have regular initial data at $t=0$ and give regular solutions for $t>0$. The allowed scale invariant initial data is just a constant, $Z_{IR}(0,x)=\Theta (x)/\sqrt{\pi}$, $Z_{UV}(0,x)=\Theta(x)/(2\sqrt{\pi})$. 
 
 Rescaling $Z(t,x)$ by such counter terms changes the diffusion equation that $Z_{\rm ren}$ solves, but it is only a sort of canonical transformation. The $Z_{\rm ren}$ functions satisfy a diffusion equation similar to that of $Z$, but with a modified potential, and $\partial_x$ replaced by ``covariant" derivatives. More generally, if the original equation is \eqref{eq:toywithpot}
and we put $Z_{\rm ren}(t,x)=e^{-S_{c.t.}(x^2/t)}Z(t,x)$, the new equation is
\beq
\partial_t Z_{\rm ren} = [\partial_x + \partial_x S_{c.t.}]^2 Z_{\rm ren}-[\frac{1}{t}V(\frac{x^2}{t})+\partial_t S_{c.t.}]Z_{\rm ren}.
\eeq
The $Z_{\rm ren}$ functions in \eqref{eq:toybranches} each solve such a modified diffusion equation on the half line\footnote{The variable $x$ would correspond to the volume density in a WDW equation, so it is natural to restrict to positive values.} $x>0$ with regular boundary condition at $t=0$. These equations are the mini superspace analogues of the $T^2$ flow equation \eqref{eq:T2flow1} of \cite{Hartman:2018tkw}.

This example illustrates that in order to translate the WDW equation into a diffusion type RG-flow equation with meaningful initial data, we are forced to introduce counter terms and consider the field theoretic flow equation of \cite{Hartman:2018tkw}. There is a different ``renormalized" flow equation for the two different branches of the solution to the WDW equation, and they differ by the choice of $S_{c.t.}(x^2/t)$. In the above example, we should think of $g(a)$ (or $\psi(a)$) as the radial WDW wave function, $Z(x,t)$ as the bare partition function, and $Z^{IR}_{\rm ren}$ as the renormalized partition function.

Turning to the full WDW equation \eqref{eq:WDW1}, Freidel shows in \cite{Freidel:2008sh} that a generic solution to this equation in Euclidean AdS has the asymptotic form
\beq
\label{eq:Freidelasymptotics}
\Psi[\gamma/\lambda^{\frac{2}{d}}] \approx e^{ S_{c.t.}}Z_+[\gamma] + e^{- S_{c.t.}}Z_-[\gamma]
\eeq
as we take $\lambda \rightarrow 0$. Here, $Z_{\pm}$ are functionals satisfying the anomalous scaling Ward identities. The possible anomaly arises because the allowed asymptotics to the WDW equation break the scaling Ward identity \eqref{eq:scaling}.

$S_{c.t.}$ is the same holographic counter term action as in \eqref{eq:ct1}\footnote{In the analysis of \eqref{eq:WDW1} following \cite{Freidel:2008sh}, $\tilde S_{c.t}$ must also include the boundary cosmological constant. It is understood that after we do the canonical transformation \eqref{eq:cantraf}, the WDW equation is \eqref{eq:WDW0}, we no longer include the boundary cosmological constant in $\tilde S_{c.t}$. The price is that the two asymptotics in \eqref{eq:Freidelasymptotics} are no longer the inverse of each other. See the transition between $g(a)$ and $\psi(a)$ in the mini-superspace example.}
\beq
 S_{c.t.} = {\tilde S}_{c.t.}[\gamma/\lambda^{\frac{2}{d}}] + \frac{1}{d} \int \log \lambda\sqrt{\gamma} A_d[\gamma],
\eeq
where $\tilde S_{c.t.}$ is the counter term action that only depends on curvature invariants of its single argument, while $A_d$ is the local conformal anomaly.
The role of the anomaly piece is to make sure that the bare partition function is invariant under the operation \eqref{eq:scaling}. 
This counter term action is the same as the one derived in \cite{Henningson:1998gx,deHaro:2000vlm}, but we emphasize that \cite{Freidel:2008sh} obtains this by analysing the asymptotic form of the solutions of the WDW equation for large scale factor, which does not require fixing a gauge.

\subsection{Diffusion kernel for the $T^2$ flow without potential}\label{sec:nopotkernel}

\subsubsection*{Hubbard-Stratonovich representation}

Let us first discuss a simplified flow equation that we get from \eqref{eq:T2flow1} by omitting the Ricci potential:
 \beq
 \label{eq:T2flow_noricc}
 \frac{\delta}{\delta \lambda} Z^{(\lambda)}(\gamma) = \frac{4}{\sqrt{\gamma}}G^{(\gamma)}_{ijkl}\frac{\delta}{\delta \gamma_{ij}}\frac{\delta}{\delta \gamma_{kl}}Z^{(\lambda)}(\gamma)  .
 \eeq
 As shown in \cite{Hartman:2018tkw} (and \cite{McGough:2016lol,Cardy:2018sdv} in the 2d case), the deformation by the $G^{(\gamma)}_{ijkl}T^{ij}T^{kl}$ operator can be represented as a coupling to a random background metric, via a Hubbard-Stratonovich (HS) transformation. The precise way to think about this representation is that it gives the evolution of $Z^{(\lambda)}$ under an infinitesimal change in $\lambda$. That is, \eqref{eq:T2flow_noricc} is equivalent with the recursive rule\footnote{Note that the covariance matrix of the Gaussian bellow is not positive definite. This is related to the usual conformal mode problem of the Euclidean gravitational path integral. We will not analyse this problem in this paper, just simply assume that one can pick complex contours such that the integrals can be evaluated.}
 \bea
\label{eq:inttransf}
Z^{(\lambda+\delta \lambda)}[\gamma]&=\frac{1}{\mathcal{N}[\gamma]}\int \mathcal{D} h e^{\frac{1}{16 \delta \lambda} \int d^dx \sqrt{\gamma}(h^2-h_{ij} h^{ij})}Z^{(\lambda)}[\gamma+h],\\
\mathcal{N}[\gamma]&=\int \mathcal{D} h e^{\frac{1}{16 \delta \lambda} \int d^dx \sqrt{\gamma}(h^2-h_{ij} h^{ij})}.
\eea
The factor of $\mathcal{N}[\gamma]$ is necessary to make sure that the l.h.s. and the r.h.s. agree in the $\delta \lambda \rightarrow 0$ limit. The $h_{ij}$ is the random HS metric. 

Let us first show that \eqref{eq:inttransf} is indeed equivalent to \eqref{eq:T2flow_noricc}. To obtain $ \frac{\delta}{\delta \lambda} Z^{(\lambda)}$, we need to take $\delta\lambda \rightarrow 0$ limit in \eqref{eq:inttransf}. In this limit, we can evaluate the $h$ integral by saddle point. It turns out that to evaluate the $\lambda$ derivative exactly, we need to care about the one loop determinant. Let us spell this calculation out in a little more detail because of this subtlety. We start by rescaling the integration variable $h\rightarrow 4 \delta \lambda h$ in \eqref{eq:inttransf} and write it as
\beq
\frac{Z^{(\lambda+\delta \lambda)}[\gamma]}{Z^{(\lambda)}[\gamma]} = \frac{\int \mathcal{D} h e^{-\delta \lambda h \cdot \Gamma \cdot h +  \delta \lambda S_1 \cdot h+ \frac{\delta \lambda^2}{2} h \cdot S_2 \cdot h + O(\delta \lambda^3)}}{\int \mathcal{D} h e^{-\delta \lambda h \cdot \Gamma \cdot h }}.
\eeq
Here, $\cdot$ refers to integral kernel plus matrix index product, and we have defined the shorthand
\bea
\Gamma & = \delta(x-y)\sqrt{\gamma}\left[\frac{1}{2}(\gamma^{ik}\gamma^{jl}+\gamma^{jk}\gamma^{il})-\gamma^{ij}\gamma^{kl}\right],\\
S_1 &= 4\frac{\delta \log Z^{(\lambda)}}{\delta \gamma_{ij}} = 2\sqrt{\gamma}\langle T^{ij} \rangle,\\
S_2 &= 16\frac{\delta^2 \log Z^{(\lambda)}}{\delta \gamma_{ij}(x)\delta \gamma_{kl}(y)}=4\sqrt{\gamma_x}\sqrt{\gamma_y}\langle T^{ij}(x)T^{kl}(y)\rangle_{\gamma,{\rm conn}}.
\eea
We can evaluate the Gaussian approximation to the integral as $\delta \lambda \rightarrow 0$ as
\beq
\frac{Z_{QFT}^{(\lambda+\delta \lambda)}[\gamma]}{Z_{QFT}^{(\lambda)}[\gamma]}= \sqrt{\frac{\det\Gamma}{\det(\Gamma-\frac{\delta\lambda}{2}S_2)}} \exp[\frac{1}{4}\delta \lambda S_1 \cdot \Gamma^{-1} \cdot S_1 + O(\delta \lambda^2)]
\eeq
The subtlety is that while $S_2$ only contributes to quadratic order in $\delta \lambda$ in the exponent, it has a linear order contribution from the one loop determinant, which we can expand as
\beq
\sqrt{\frac{\det\Gamma}{\det(\Gamma-\frac{\delta\lambda}{2}S_2)}} \approx 1+\frac{1}{4}\delta\lambda \text{Tr}(\Gamma^{-1}\cdot S_2) + O(\delta \lambda^2),
\eeq
which leads to the result
\bea
\frac{Z_{QFT}^{(\lambda+\delta \lambda)}[\gamma]}{Z_{QFT}^{(\lambda)}[\gamma]} &=  1+\frac{\delta \lambda}{4}[S_1 \cdot \Gamma^{-1} \cdot S_1 + \text{Tr}(\Gamma^{-1} \cdot S_2)]+O(\delta \lambda^2) \\
&= 1+\delta \lambda \int d^d x \sqrt{\gamma}(\gamma_{ik}\gamma_{jl}-\frac{1}{d-1}\gamma_{ij}\gamma_{kl})\langle T^{ij}(x)T^{kl}(x)\rangle+ O(\delta \lambda^2),
\eea
where we have used that $\Gamma^{-1}=\frac{1}{\sqrt{\gamma}}G^{(\gamma)}_{ijkl}\delta(x-y)=\frac{1}{\sqrt{\gamma}}[\frac{1}{2}(\gamma_{ik}\gamma_{jl}+\gamma_{jk}\gamma_{il})-\frac{1}{d-1}\gamma_{ij}\gamma_{kl}]\delta(x-y)$.
The two point function entering is the non-connected one, the effect of the $S_1^2$ terms is to cancel the disconnected piece from $S_2 \sim \langle T^2 \rangle - \langle T \rangle^2$. This leads to the flow equation 
\beq
\label{eq:flow}
\frac{d}{d\lambda} \log Z_{QFT}^{(\lambda)}[\gamma] =  \int d^d x \sqrt{\gamma}(\gamma_{ik}\gamma_{jl}-\frac{1}{d-1}\gamma_{ij}\gamma_{kl})\langle T^{ij}(x)T^{kl}(x)\rangle.
\eeq
which is equivalent to \eqref{eq:T2flow_noricc}.

\subsubsection*{Iterating the kernel: constant $\lambda$}

Given the recursive relation \eqref{eq:inttransf}, we can formally construct the diffusion kernel, by simply iterating it.\footnote{Related ideas and calculations can be found in \cite{Lee:2012xba,Lee:2013dln,Mazenc:2019cfg,Kim:2020nst,Kim:2020kma}.} For simplicity, let us first assume that $\lambda$ is a constant, and we iterate with the same constant $\delta \lambda$ steps. That is, instead of \eqref{eq:T2flow_noricc}, we are solving the integrated version of the equation
\beq
 \frac{d}{d \lambda} Z^{(\lambda)}(\gamma) = \int d^d x\frac{4}{\sqrt{\gamma}}G^{(\gamma)}_{ijkl}\frac{\delta}{\delta \gamma_{ij}}\frac{\delta}{\delta \gamma_{kl}}Z^{(\lambda)}(\gamma)  .
\eeq
Let us set $\delta \lambda= \lambda/N$ for some large number $N$. We can iterate \eqref{eq:inttransf} $N$ times to get 
\bea
Z^{(\lambda)}[\gamma]&=\int \prod_{k=1}^N \frac{\mathcal{D} h_k}{\mathcal{N}(\gamma+\sum_{l=0}^{k-1} h_l)} \\ &\times \exp \left({\frac{1}{16}\frac{N}{\lambda} \sum_{k=1}^N \int d^d x \sqrt{|\gamma+\sum_{l=0}^{k-1} h_l|}[h_k^2-(h_k)_{ij}(h_k)^{ij}]}\right) Z^{(0)} [\gamma + \sum_k^N h_k].
\eea
The $Z^{(0)}$ is the initial data of the flow. As we have discussed in the previous section, there will be no healthy initial data for this simplified flow that is consistent with the scaling property \eqref{eq:scaling}. Such initial data only exists for the renormalized flow equation, so we imagine for now that we stop at some finite small $\lambda$.
We then define the rescaled integral variable via
\beq
(h_k)_{ij}=\frac{\lambda}{N} H_{ij}(\lambda \frac{k}{N}).
\eeq
Note that we interpret $H_{ij}$ as having an extra variable $\eta \equiv \lambda \frac{k}{N}$, hence living in one dimension higher than $h_{ij}$. With this reparametrization, all the scalings work out nicely and we can turn all the sums into integrals in the $N\rightarrow \infty$ limit, giving
\bea
Z^{(\lambda)}[\gamma]&=\int \prod_{0<\eta<\lambda}\frac{ \mathcal{D} H(\eta)}{\mathcal{N}(\gamma+\int_0^\eta d\mu K(\mu))}\\
&\times \exp \left({\frac{1}{16}\int_0^\lambda d\eta \int d^d x \sqrt{|\gamma+\int_0^\eta d\mu H(\mu)|}[H(\eta)^2-H(\eta)_{ij}H(\eta)^{ij}]}\right) \\ &\times Z^{(0)} [\gamma + \int_0^\lambda d\eta H(\eta)].
\eea
The final step is to change variable in the path integral to
\beq
 \gamma(\eta)_{ij}=\gamma_{ij}+\int_0^\eta d\mu H(\mu)_{ij}.
\eeq
We have
\beq
H(\eta)_{ij}= \partial_\eta  \gamma(\eta)_{ij},
\eeq
and the Jacobian is
\beq
\frac{\delta  \gamma(\eta)_{ij}}{\delta H(\mu)_{kl}} = \delta_{ik}\delta_{jl} \theta(\eta-\mu),
\eeq
which is field independent, so there is no Jacobi determinant in the path integral measure. This way, we arrive at
\bea
\label{eq:iteration1}
Z^{(\lambda)}[\gamma] &=\int_{ \gamma(0)=\gamma} \prod_{0\leq \eta\leq \lambda}\frac{ \mathcal{D}  \gamma(\eta)}{\mathcal{N}[ \gamma(\eta)] }\\ &\times \exp \left({\frac{1}{16}\int_0^\lambda d\eta \int d^d x \sqrt{| \gamma(\eta)|}[(\partial_\eta  \gamma (\eta))^2-\partial_\eta  \gamma(\eta)_{ij}\partial_\eta  \gamma(\eta)^{ij}]}\right) Z^{(0)} [ \gamma(\lambda)]\\
&=\int \mathcal{D} \gamma_0 L[\gamma,\gamma_0]Z^{(0)}[\gamma_0],
\eea
where in the last line we have defined the kernel
\bea
\label{eq:WDWpropagator}
L[\gamma,\gamma_0] &= \int_{\gamma(0)=\gamma,  \gamma(\lambda)=\gamma_0} \prod_{0\leq \eta \leq \lambda} \frac{\mathcal{D}  \gamma(\eta)}{{\mathcal{N}[ \gamma(\eta)] }}\\ &\times \exp \left({\frac{1}{16}\int_0^\lambda d\eta \int d^d x \sqrt{| \gamma(\eta)|}[(\partial_\eta  \gamma (\eta))^2-\partial_\eta  \gamma(\eta)_{ij}\partial_\eta  \gamma(\eta)^{ij}]}\right) .
\eea
Notice that the combination in the exponential is $K^2-K_{ij}K^{ij}$, where $K_{ij}$ is the extrinsic curvature for the constant $\eta$ slices of a geometry $d\eta^2+ \gamma_{ij}(\eta) dx^i dx^j$. This is a path integral in $d+1$ dimension, but not a covariant one: it only includes the kinetic terms of the Einstein action. To obtain a gravitational path integral, we need to include the Ricci potential as in \eqref{eq:T2flow1}, to which we turn in sec. \ref{sec:addingRicci}. This potential will be fixed by requiring a consistent solution to \eqref{eq:T2flow1} in the case where $\lambda=\lambda(x)$ is allowed to depend on the $x$ coordinates.

\subsubsection*{Iterating the kernel: position dependent $\lambda$}

We can repeat the above exercise when $\lambda(x)$ depends on the $d$-dimensional coordinates $x$, and in this case we can choose an iteration scheme, where each step is a different function $\delta \lambda_k(x)$ such that
\beq
\lambda(x)=\sum_{k=1}^N \delta \lambda_k(x).
\eeq
We want to turn sums into integrals as before, therefore let us introduce the variable $\eta=k/N$, $d\eta = 1/N$, $\eta \in [0,1]$, such that
\bea
\delta \lambda_k(x) &= \frac{1}{N}\mu(\frac{k}{N},x) \equiv d\eta \mu(\eta,x),\\
\lambda(x) &=\int^1_0 d\eta \mu(\eta,x),
\eea
where $\mu(\eta,x)$ is now a $d+1$ dimensional function parametrizing the path in $\lambda$ space. As before, we first change variable to $H_{ij}$ via
\beq
(h_k)_{ij}=\delta \lambda_k H_{ij}(\frac{k}{N},x) \equiv d\eta \mu(\eta,x) H_{ij}(\eta,x),
\eeq
and then to
\beq
\label{eq:ZmetricCont}
 \gamma_{ij}(\eta,x)=\gamma_{ij}(x)+ \int_0^\eta d\eta' \mu(\eta',x)H_{ij}(\eta',x).
\eeq
One can easily check that while the independent Jacobians depend on $\mu(\eta,x)$, the combined Jacobian is independent of it. Repeating the manipulations leading to \eqref{eq:iteration1}, we arrive at
\bea
\label{eq:iteration2}
Z&^{(\lambda(x))}[\gamma] =\int_{\gamma(0)=\gamma} \prod_{0\leq \eta\leq 1}\frac{ \mathcal{D}  \gamma(\eta)}{\mathcal{N}[\mu(\eta), \gamma(\eta)] }\\ &\times \exp \left({\frac{1}{16}\int_0^1 d\eta \int d^d x \frac{\sqrt{| \gamma(\eta)|}}{\mu(\eta,x)}[(\partial_\eta  \gamma (\eta))^2-\partial_\eta  \gamma(\eta)_{ij}\partial_\eta  \gamma(\eta)^{ij}]}\right) Z^{(0)} [ \gamma(1)]\\
&=\int \mathcal{D} \gamma_0 L[\mu,\gamma,\gamma_0]Z^{(0)}[\gamma_0].
\eea
This new kernel $L[\mu,\gamma,\gamma_0]$ depends on $\mu(\eta,x)$, i.e. the path that we took to build the iteration. This is a highly undesirable situation, since it should only depend on the endpoints of the path 
\beq
\label{eq:path}
\lambda(\eta,x)=\int_0^\eta d\eta' \mu(\eta',x),
\eeq
that we are taking in coupling space. This path dependence is due to the lack of $d+1$ dimensional diffeomorphism invariance in the path integral, and is related to the failure of \eqref{eq:T2flow_noricc} to satisfy certain Wess-Zumino consistency conditions, see e.g. \cite{Shyam:2016zuk,Shyam:2017qlr}. We will see in the following section that adding the Ricci potential as in \eqref{eq:T2flow1} with a specific coefficient solves part of the problem, namely it turns the exponential into a path-independent diff-invariant object.

Notice however that there is also a residual dependence on the field $\mu(\eta,x)$ in the measure factor $\mathcal{N}[\mu(\eta), \gamma(\eta)]$. This dependence comes directly from writing $\delta \lambda=d\eta \mu$ in the definition \eqref{eq:inttransf} and it will turn out to be crucial as we explain in section \ref{sec:measure}.

For the remainder of this paper, we will assume we are working with a spatially varying coupling $\lambda(x)$, but we will often not write the $x$-dependence explicitly to avoid cluttering expressions. We will also sometimes use $\lambda$ for the path in coupling space $\lambda(\eta,x)$ of \eqref{eq:path}, which should be clear from context.

\subsection{Diffusion kernel for the $T^2$ flow with Ricci potential}
\label{sec:addingRicci}

We will now show that there is a simple way to restore path-independence in the kernel derived in the previous section. This section will focus on the exponential and we will see that adding the appropriate potential to the diffusion equation will restore diffeomorphism inviarance in an elegant way: the kernel will simply become the path-integral with the $d+1$-dimensional Einstein-Hilbert action.

We want to iterate the flow equation \rref{eq:T2flow1}. To do so, we consider the slight modification of the Hubbard-Strotonovich trick 
\be
Z^{(\lambda+\delta \lambda)}[\gamma]=\frac{e^{\int d^d x\delta\lambda V(\lambda,\gamma)}}{\mathcal{N}[\gamma]}\int \mathcal{D} h e^{ \int d^dx \frac{1}{16 \delta \lambda} \sqrt{\gamma}(h^2-h_{ij} h^{ij})}Z^{(\lambda)}[\gamma+h] \,.
\ee
Repeating the steps after \eqref{eq:inttransf}, we find that this recursion relation is equivalent to the flow equation
 \beq
 \label{eq:flowwithgeneralpot}
 \frac{\delta}{\delta \lambda} Z^{(\lambda)}(\gamma) = \frac{
 4}{\sqrt{\gamma}}G^{(\gamma)}_{ijkl}\frac{\delta}{\delta \gamma_{ij}}\frac{\delta}{\delta \gamma_{kl}}Z^{(\lambda)}(\gamma)+ V(\lambda,\gamma) Z^{(\lambda)}(\gamma)  .
 \eeq
For this potential to preserve the scaling symmetry \eqref{eq:scaling} of the flow equation, we must demand that
\be
\label{eq:potentialscaling}
V(\alpha(x)^{d} \lambda, \alpha(x)^2 \gamma)=\alpha(x)^{-d} V(\lambda, \gamma) \,.
\ee
The following Ricci potential preserves this symmetry
\be
V(\lambda,\gamma)= a_d \lambda^{-1} \sqrt{\gamma/\lambda^{2/d}} R_{\gamma/\lambda^{2/d}}
\ee
where the prefactor of $\lambda^{-1}$ was chosen precisely for this purpose. Adding this Ricci potential clearly has no impact on the path-integral measure, so the only thing left to do is track its effect on the exponential as we iterate. It is convenient in what follows to work with what will become an induced metric
\be
q_{ij}(x)=\gamma_{ij}(x)/\lambda(x)^{2/d} \,.
\ee
The iteration yields the following contribution to the action
\be
a_{d}\int d\eta  \p_\eta \log \lambda(\eta,x) \int d^dx\sqrt{q(\eta)} R_{q(\eta)} \,,
\ee
where
\be \label{inducedmetric}
q_{ij}(\eta)= \frac{\gamma_{ij}+\int_0^\eta d\eta' \mu(\eta')H_{ij}(\eta')}{(\lambda_0+\int_0^\eta d\eta'\mu(\eta'))^{2/d}}
\ee
We will now see that this term can be recombined with the rest of the action into the Einstein-Hilbert action, up to terms that will have a nice interpretation. This will also make the action manifestly path-independent. To see this, consider the following metric ansatz
\be
\label{eq:bulkmetric}
ds^2= N^2(\eta,x) d\eta^2+\Omega^2(\eta,x) \gamma_{ij}(\eta,x)dx^idx^j \equiv g_{ab}dx^a dx^b\,.
\ee
With this gauge choice, we have on a fixed $\eta$ surface 
\be
\sqrt{g}(K^2-K^{ij}K_{ij})=\frac{\sqrt{\gamma}}{N}\left(d(d-1)\Omega^{d-2}(\dot{\Omega})^2+(d-1)\Omega^{d-1}\dot{\Omega}\dot{\gamma}+\frac{\Omega^d}{4}\left(\dot{\gamma}^2-\dot{\gamma}^{ij}\dot{\gamma}_{ij}\right)\right) \,,
\ee
where $K_{ij}$ is the extrinsic curvature tensor. We can use this relation to eliminate $\dot{\gamma}^{ij}$ in \eqref{eq:iteration2}. We therefore obtain the total action to be
\be \label{action}
\int d\eta \int d^d x \frac{\sqrt{g}}{\mu}[\frac{N}{4\Omega^{d}}(K^2-K^{ij}K_{ij})]+a_d\int d\eta\p_\eta \log \lambda  \int d^d x  \sqrt{\gamma/\lambda^{2/d}} R_{\gamma/\lambda^{2/d}}+S_{\text{rest}} \,,
\ee
with
\bea
S_{\text{rest}}&=-\frac{1}{4}\int d\eta \int d^dx \frac{\sqrt{\gamma}}{\mu}\left(d(d-1)\frac{(\dot\Omega)^2}{\Omega^2}+(d-1)\frac{\dot\Omega \dot \gamma}{\Omega}\right) \notag \\
&=-\frac{1}{4}\int d\eta \int d^dx \frac{\sqrt{\gamma}}{\mu}\left(d(d-1)\frac{(\dot\Omega)^2}{\Omega^2}+2(d-1)\frac{\dot\Omega \p_\eta\sqrt{\gamma}}{\Omega \sqrt{\gamma}}\right)
\eea

Analyzing \rref{action}, demanding that the $\lambda$ and $\mu$ dependence disappears from the action leads to the following parametrization
\be \label{parametrization}
\Omega(\eta,x) = b \lambda(\mu,x)^{-1/d} \,, \qquad N(\eta,x) = c \mu(\eta,x) \lambda(\eta,x)^{-1} \,,
\ee
where $b,c$ are constants that are undetermined at this point. Note that this way, the induced metric on a constant $\eta$ surface is simply $b^2q$ with $q$ defined in \rref{inducedmetric}. With this choice of parametrization, the action then becomes
\be
\frac{c}{4b^d}\int d\eta \int d^d x \sqrt{g}(K^2-K^{ij}K_{ij})+\frac{a_d}{b^{d-2}c}\int d\eta \int d^d x  \sqrt{g} R_{b^2q}+S_{\text{rest}}
\ee
Note that each of the first two terms are separately invariant under diffeomorphism of the form $\eta\to f(\eta)$, but they are not invariant under diffeomorphism of the form $\eta\to f(\eta,x)$. To restore full $d+1$-dimensional diff-invariance we require a non-zero $a_d$ which satisfies $\frac{c}{4b^d}=\frac{a_d}{b^{d-2}c}$. We will denote $\frac{c}{4b^d}=\xi$, which together with the Gauss-Codazzi equation, gives us\footnote{More precisely, we are using the relation
\beq
R_g=R_{b^2q}+K^2-K^{ij}K_{ij}-\partial_\eta \big[ 2\sqrt{|b^2 q|}K\big]-\frac{2}{N}\nabla^i\nabla_i N,
\eeq
which can be derived from the Gauss-Codazzi equations for metrics in the gauge \eqref{eq:bulkmetric}. After using this identity, we have dropped the term $\int d\eta \left[ \int d^d x \sqrt{|b^2 q|}2 \nabla^i \partial_i N \right]$ from the action because it is a boundary term on the transverse manifold where the field theory is defined. We are interested in two situations, one where this manifold is compact (corresponding to usual radial flow), and the other is when we keep the flow parameter $\lambda=0$ at the boundary of this manifold (corresponding to flowing e.g. in Euclidean York time). In both cases, dropping this term is justified.}
\be
S=\xi \int d\eta \int d^dx \sqrt{g}R_g + 2 \xi \int d^d x \sqrt{b^2 q} K \Bigg|_{\text{endpoints}} + S_{\text{rest}} \,.
\ee

We have thus recovered the Einstein-Hilbert action with the Gibbons-Hawking term, for now without the cosmological constant. We will see that in fact it comes from $S_{\text{rest}}$. Plugging \rref{parametrization} into the definition of $S_{\text{rest}}$, we find
\begin{eqnarray}
S_{\text{rest}}&=&\frac{1}{4b^d c}\frac{d-1}{d}\int d\eta \int d^dx \sqrt{g}+\frac{1}{2b^d }\frac{d-1}{d}\int d^dx \sqrt{b^2 q} \Bigg|_{\text{endpoints}} \notag \\
&=& \frac{\xi}{c^2}\frac{d-1}{d}\int d\eta \int d^dx \sqrt{g}+\frac{2\xi}{c}\frac{d-1}{d}\int d^dx \sqrt{b^2 q} \Bigg|_{\text{endpoints}} \,.
\end{eqnarray}
We can now set $c=\frac{\ell_{\text{AdS}}}{d}$ and $\xi=1$ which fixes all the free parameters. Note that this precisely matches the choice of $a_d$ obtained below \rref{eq:WDW1}. Reinstating the factor of $16\pi G_N$, we obtain
\be
S_{\text{tot}}=\frac{1}{16\pi G_N} \int d\eta d^dx \sqrt{g}(R-2\Lambda) + \left[ \frac{1}{8\pi G_N} \int d^d x\sqrt{b^2 q} (K+ \frac{d-1}{\ell_{\text{AdS}}})\right]_{\text{endpoints}} \,,
\ee
namely the Euclidean Einstein-Hilbert action with cosmological constant, along with the Gibbons-Hawking term and the leading holographic counterterm. The overall sign of the action is the opposite of the usual Euclidean path integral whose weight is $e^{-S_E}$ and we will comment on this fact below. Also note that we obtain the action with the right prefactor due to our choice of $\xi$, which was an overall free-parameter and not imposed on us by first principles. The choice of $\xi$ enters only in the dictionary between the bulk induced metric and the QFT background metric.

The formal solution of the flow equation \eqref{eq:T2flow1} is therefore
\bea
\label{eq:iteration3}
Z^{(\lambda(x))}[\gamma] &=\int_{q(0)=\gamma/\lambda^{2/d} } \mathcal{D}M(q)e^{S_{\rm tot}[g]} Z^{(\lambda'(x))} [{\lambda'}^{2/d}q(1)]\\
&\equiv \int \mathcal{D} \gamma_0 L[\mu,\gamma/\lambda^{2/d},\gamma_0/{\lambda'}^{2/d}]Z^{(\lambda'(x))}[\gamma_0],
\eea
where the equation defines the final diffusion kernel, which is the Euclidean gravitational path integral with Dirichlet condition between two boundaries. 

As explained above, we obtain the gravitational action with an overall minus sign. While this may seem surprising at first sight, it is in fact expected. The CFT ($\lambda=0$) partition function matches the bulk path integral weighted by $e^{-S_E}$. To move into bulk, we need to ``undo" the path integral between the conformal boundary and the bulk radial slice, which is accomplished by path-integrating with the opposite sign (on some complex contour for convergence). This is exactly what our kernel accomplishes. A similar observation was made in \cite{Heemskerk:2010hk}.

So far, we have left the resulting measure $\mathcal{D}M(q)$ unspecified, we analyse it further in sec. \ref{sec:measure}. Ideally, one would hope that it is a gauge fixed version of a diffeomorphism invariant measure for the $d+1$ metric $g_{ab}$. This would be required for the kernel $L$ to be independent of the iteration path $\mu(\eta,x)$. We left the $\mu$ dependence explicit in the kernel for now.

\subsubsection*{The kernel for the renormalized partition function}

Finally, let us comment on the diffusion kernel for the renormalized partition function. As explained in sec. \ref{sec:toyflows}, one may obtain it by conjugating with the canonical transformation introducing the counter terms, which is given by
\beq
L_{\rm ren}[q,q']=e^{-S_{c.t}[\lambda,q]} L[q,q']e^{S_{c.t}[\lambda',q']},
\eeq
where $S_{c.t.}$ was discussed around \eqref{eq:ct1} and we remind the reader that $q_{ij}\propto \gamma_{ij}/\lambda^{2/d}$. The explicit $\lambda$ dependence enters only in even dimensions due to the conformal anomaly.

In summary, in this section we showed that adding a Ricci-potential to the flow equation restores diffeomorphism-invariance in the action and the kernel, which then no longer depends on the choice of path. With this parametrization, we also obtain the leading holographic counter-term with the right coefficient. This concludes the section and we now turn to the path integral measure.

\subsection{Gauge invariance and the path integral measure \label{sec:measure}}

We have seen that the iteration procedure produces the Einstein-Hilbert action in the diffusion kernel solving the $T^2$ flow equation. However, we cannot claim that the kernel is a $d+1$ dimensional diffeomorphism invariant path integral, since we only integrate over the spatial metric $q_{ij}$ in the special gauge \eqref{eq:bulkmetric}. Our aim here will be to argue that this is indeed a gauge fixed version of the $d+1$ dimensional path integral. This involves examining the formal path integral measure that is induced by the iteration which will yield a small surprise.

Retracing our iteration steps deriving the action for the kernel, we find that the path integral measure we get is
\beq
\mathcal{D}M(q)=\prod_{\eta}\frac{\prod_{x,i\leq j}d\gamma_{ij}}{\int \prod_{x,i\leq j}dh_{ij} e^{-\int\frac{1}{16 d\eta \mu(\eta,x)} h \Gamma_\gamma h }}
\eeq
As mentioned during the derivation, the Jacobians arising when changing integration variable to $\gamma$ are field independent. Now we change to the variable $q=\gamma/\lambda^{2/d}$. Under this transformation, $\Gamma_\gamma=\lambda^{1-4/d}\Gamma_q$. We change variable in the denominator $\hat h=\lambda^{1/2-2/d}h/\sqrt{\mu}$ to absorb this change. The result is
\beq
\prod_{\eta}\frac{\prod_{x,i\leq j}d\gamma_{ij}}{\int \prod_{x,i\leq j}dh_{ij} e^{-\int\frac{1}{16 d\eta \mu(\eta,x)}h \Gamma_\gamma h }} \sim \prod_{x,\eta} \sqrt{\det \Gamma_q} \left(
\prod_{i\leq j}d\left[\sqrt{\frac{\lambda}{\mu}}q_{ij} \right] \right),
\eeq
where we have dropped a power of $d\eta$. We recognise the lapse $N=\mu/\lambda$. Pulling it out from the product over independent components gives
\beq
\label{eq:measurewithq}
\mathcal{D}M(q)= \prod_{x,\eta}  \sqrt{\det \Gamma_q}\left( N^{-\frac{d(d+1)}{4}}
\prod_{i\leq j}dq_{ij}  \right).
\eeq
Now we proceed to interpret this measure. We start with the most obvious interpretation which will turn out to fail.

\subsubsection*{Geometric measure}

A very natural measure \cite{Polyakov:1981rd,Mazur:1989by}, also used on the string worldsheet,  comes from considering an invariant inner product on metric variations $(\delta g,\delta g)=\delta g \cdot \Gamma_g \cdot \delta g$, where $\Gamma_g = \delta(x-y)\sqrt{g}(2g^{ab}g^{cd}-g^{ac}g^{bd}-g^{bc}g^{ad})/2$ as before. As in usual finite dimensional geometry, an invariant measure of integration is obtained by considering the normalized top form in this inner product, that is
\beq
\label{eq:PolyakovdeWitt}
\prod_{x,\eta}\sqrt{\det \Gamma_g}\prod_{a\leq b}dg_{ab}.
\eeq
The first thought would be to try to interpret \eqref{eq:measurewithq} as a gauge fixed version of this measure. Let us try to do this.
We would like to replace the integral over $q_{ij}$ by the total $d+1$ metric $g_{ab}$. We may write the measure \eqref{eq:measurewithq} as
\bea
\prod_{x,\eta}  &\sqrt{\det \Gamma_q}\left( N^{-\frac{d(d+1)}{4}}
\prod_{i\leq j}dq_{ij}  \right)= \\ &= \prod_{x,\eta} \left( \prod_{a\leq b}dg_{ab}\right) \sqrt{\det \Gamma_g} \left[ N^{-\frac{d(d+1)}{4}}\frac{\sqrt{\det \Gamma_q}}{\sqrt{\det \Gamma_g}} \prod_i\delta(g_{i\eta})\delta(g_{\eta\eta}-N^2)\right]
\eea
In order to reproduce \eqref{eq:PolyakovdeWitt}, we need the expression in the big brackets to be the product of a gauge fixing functional (the delta functions) and the corresponding Faddeev-Popov determinant \cite{Faddeev:1967fc} (the rest). In other words, we would need to show that $\Delta_{FP}=N^{-\frac{d(d+1)}{4}}\frac{\sqrt{\det \Gamma_q}}{\sqrt{\det \Gamma_g}}$ for the choice of gauge fixing.

Now one can check that
\beq
\sqrt{\det \Gamma_q} \propto (\sqrt{q})^{(d+1)(d/4-1)}.
\eeq
One way to see this is to note that due to diff invariance properties $\det \Gamma_q$ must be proportional to some power of $\det q$, and then work out how $\det \Gamma_q$ changes under rescaling $q_{ab}$ using the Gaussian integral representation of the determinant. Since in our chosen gauge $g_{ij}\equiv q_{ij}$ and $\sqrt{g}=N \sqrt{q}$, one ends up with
\beq
N^{-\frac{d(d+1)}{4}}\frac{\sqrt{\det \Gamma_q}}{\sqrt{\det \Gamma_g}} \propto N^{\frac{3-d^2}{2}} (\sqrt{q})^{\frac{1-d}{2}}.
\eeq
This is impossible to get from the FP determinant that is a determinant of a $(d+1)\times (d+1)$ matrix. The main problem is the scaling of $N$, we will never get a power that goes like $d^2$. We thus conclude that the measure we have obtained cannot be viewed as a gauge-fixed version of the geometric measure.

\subsubsection*{Interpretation as Hamiltonian path integral}

On the other hand, \eqref{eq:measurewithq} has a natural interpretation as a measure in a Hamiltonian path integral. We may write it as
\beq
\label{eq:measureq2}
\mathcal{D}M(q)=\prod_{x,\eta}  \sqrt{\det \frac{\Gamma_q}{N}}\left( 
\prod_{i\leq j}dq_{ij}  \right).
\eeq
The idea is to interpret the determinant as coming from a Gaussian integral with a kernel $N\Gamma_q^{-1}$ where $\Gamma_q^{-1}=\frac{1}{2\sqrt{q}}[\frac{2}{d-1}q_{ij}q_{kl}-q_{ik}q_{jl}-q_{jk}q_{il}]\delta(x-y)$. Such a new integration variable can be interpreted as a gravitational canonical momentum, as we will now explain. In the ADM decomposition \cite{Arnowitt:1959ah}, the bulk part of the Lagrangian can be written as
\beq
\label{eq:ADMLagrangian}
\mathcal{L}=P^{ij}\partial_\eta q_{ij}+N^i \mathcal{H}_i + N \mathcal{H},
\eeq
where for us $N^i=0$ and $P^{ij}$ is the canonical momentum. The Hamiltonian constraint contains the $P$ dependence
\beq
\int d^d x N \mathcal{H} =  P \cdot N(\Gamma_q)^{-1} \cdot P + \text{terms only depending on }q_{ij}
\eeq
and the canonical momentum $P^{ij}$ is determined in terms of $q_{ij}$ by solving the equations of motion coming from variations of \eqref{eq:ADMLagrangian} with respect to $P^{ij}$. Now since the action is Gaussian in $P^{ij}$, we may equivalently write the path integral as having an extra independent integral over the $P^{ij}$, and this Gaussian integral enforces the equation of motion for $P$ and in addition precisely produces the prefactor in the measure \eqref{eq:measureq2} as a one loop determinant. Therefore, the (bare) diffusion kernel has the schematic form
\beq
\label{eq:admkernel}
L = \int \mathcal{D}q_{ij}\mathcal{D}P^{ij} e^{-\int(P^{ij}\partial_\eta q_{ij}+N \mathcal{H}[P,    q])+\text{bndy terms}},
\eeq
where now the measure is \textit{flat} (i.e. translationally invariant) for both $q_{ij}$ and $P^{ij}$. As argued in the previous section, this cannot be turned into the full diff invariant $d+1$ dimensional measure. However, it is part of a natural measure for the Hamiltonian path integral explored in \cite{Han:2009bb}  (see also \cite{Faddeev:1973zb,Fradkin:1974df} for earlier discussions of the Hamiltonian path integral in gravity). This reference shows that the flat measure
\beq
\mathcal{D}q_{ij}\mathcal{D}P^{ij} \mathcal{D}N^i \mathcal{D}N
\eeq
is gauge invariant under the gauge transformations generated by the first class constraints of GR via the Poisson bracket. The enveloping algebra of these constraints is called the Bergmann-Komar group, which is actually only a group on-shell, in which case it coincides with usual spacetime diffeomorphisms. Off-shell, the measure is anomalous under diffeos mixing $\eta$ and $x^i$, but instead of diffeos, the full BK-group is a non-anomalous gauge symmetry of it. 

In our case, we are missing the $N$ and $N^i$ integrals in \eqref{eq:admkernel}, so we want to interpret the integral in the kernel as being gauge fixed. As usual, the gauge fixed path integral
\beq
  \mathcal{D}q_{ij}\mathcal{D}P^{ij} \int \mathcal{D}N^i \mathcal{D}N \Delta_{FP} \delta(N^i)\delta(N-c \partial_\eta \log \lambda).
\eeq
is equivalent with the non-gauge fixed one provided the Faddeev-Popov determinant $\Delta_{FP}=\vline \frac{\delta(N,N^i)}{\delta \epsilon^a}\vline$ is included.
 It is important that the variation is with respect to the BK gauge transformations considered in \cite{Han:2009bb}, which for the lapse and shift are always related to regular diffeos $\xi^a$ via $\xi^\eta=\epsilon/N$ and $\xi^i=\epsilon^i-\epsilon N^i/N$, where $(\epsilon,\epsilon^i)$ are the parameters of the BK gauge transformation. One has \cite{Han:2009bb}
\bea
\delta N &= \epsilon^j \partial_j N-N^j \partial_j \epsilon+\partial_\eta \epsilon \\
\delta N^i &= \epsilon^k \partial_k N^i - (\partial_k \epsilon^i)N^k + q^{ik}(\epsilon\partial_k N-N\partial_k \epsilon)+\partial_\eta \epsilon^i,
\eea
leading to the Faddeev-Popov determinant
\beq
\Delta_{FP}= \; \vline 
\begin{array}{cc}
 -\partial_\eta    & (\partial_j N) \\
 q^{ik}[(\partial_k N)-N\partial_k ]    & \delta^i_j \partial_\eta
\end{array}
 \vline \;,
\eeq
where $N=c \partial_\eta \log \lambda$.
In order for the diffusion kernel $L$ for the $T^2$ flow to be given by a $d+1$ dim path integral that is gauge invariant in the sense of \cite{Han:2009bb}, one needs $\Delta_{FP}$ to be field independent. In general this does not seem to be the case, but it is the case when the lapse is independent of the $x^i$ coordinates, $\partial_i N=0$, since in this case we get a lower triangular FP matrix and $\Delta_{FP}=(\det \partial_\eta)^{d+1}$ which is field independent. Having $\partial_i N=0$ is equivalent with
\beq
\partial_i \partial_\eta \log \lambda=0 \quad \Rightarrow \quad \lambda(\eta,x)=f(\eta)\lambda_0(x).
\eeq
Therefore it seems like we cannot trace an arbitrary path in coupling space, but it is possible to start from an arbitrary profile and flow along that. This seems to be sufficient, since the profile $\lambda_0(x)$ only carries information at the conformal end of the flow; on the other end it is redundant with the scale factor $\sqrt{|q|}$.

For such paths, we may therefore write the diffusion kernel in the un-gauge fixed form schematically as
\beq
L = \int\mathcal{D}q_{ij}\mathcal{D}P^{ij} \mathcal{D}N^i \mathcal{D}N  e^{-\int(P^{ij}\partial_\eta q_{ij}+N \mathcal{H}+N^i\mathcal{H}_i)+\text{bndy terms}},
\eeq
where $\mathcal{H}_i$ are the momentum constraints.

Finally, we would like to mention that the WDW equation has ordering ambiguities which are capable of affecting the path integral measure \cite{Parker:1979mf,Halliwell:1988wc}, and the $T^2$ flow only partially resolves factor ordering. It would be interesting to understand this better.

\section{Hartle-Hawking wave-functions and the extremal volume}
\label{sec:3}

Here we will discuss some possible applications of a position dependent $T^2$ flow.

\subsection{Wave functions and metric eigenstates in AdS}

As we have reviewed, the flow equation \eqref{eq:T2flow1} originates from the WDW equation \eqref{eq:WDW0} by introducing an auxiliary scale $\lambda(x)$ such that $Z^{(\lambda)}[\gamma]=\Psi[\gamma/\lambda^{2/d}]$, where $Z$ is a bare partition function and $\Psi$ is related to a solution of the WDW equation \eqref{eq:WDW1} by a canonical transformation. It is common to think about $\Psi$ as a radial wave function. However, as explained e.g. in \cite{Freidel:2008sh,Kraus:2018xrn}, there is no built in radial coordinate in $\Psi$, the radial dependence of $\Psi$ is determined roughly by the scale factor of the metric, with $\sqrt{|\gamma/\lambda^{2/d}|}\rightarrow \infty$ or $\lambda \rightarrow 0$ corresponding to the AdS boundary. This implies that $\Psi$ can also be appropriately regarded as a bulk wavefunction describing states in the CFT. Let us outline how this interpretation works.

\begin{figure}[!h]
\begin{center}
\includegraphics[width=0.3\textwidth]{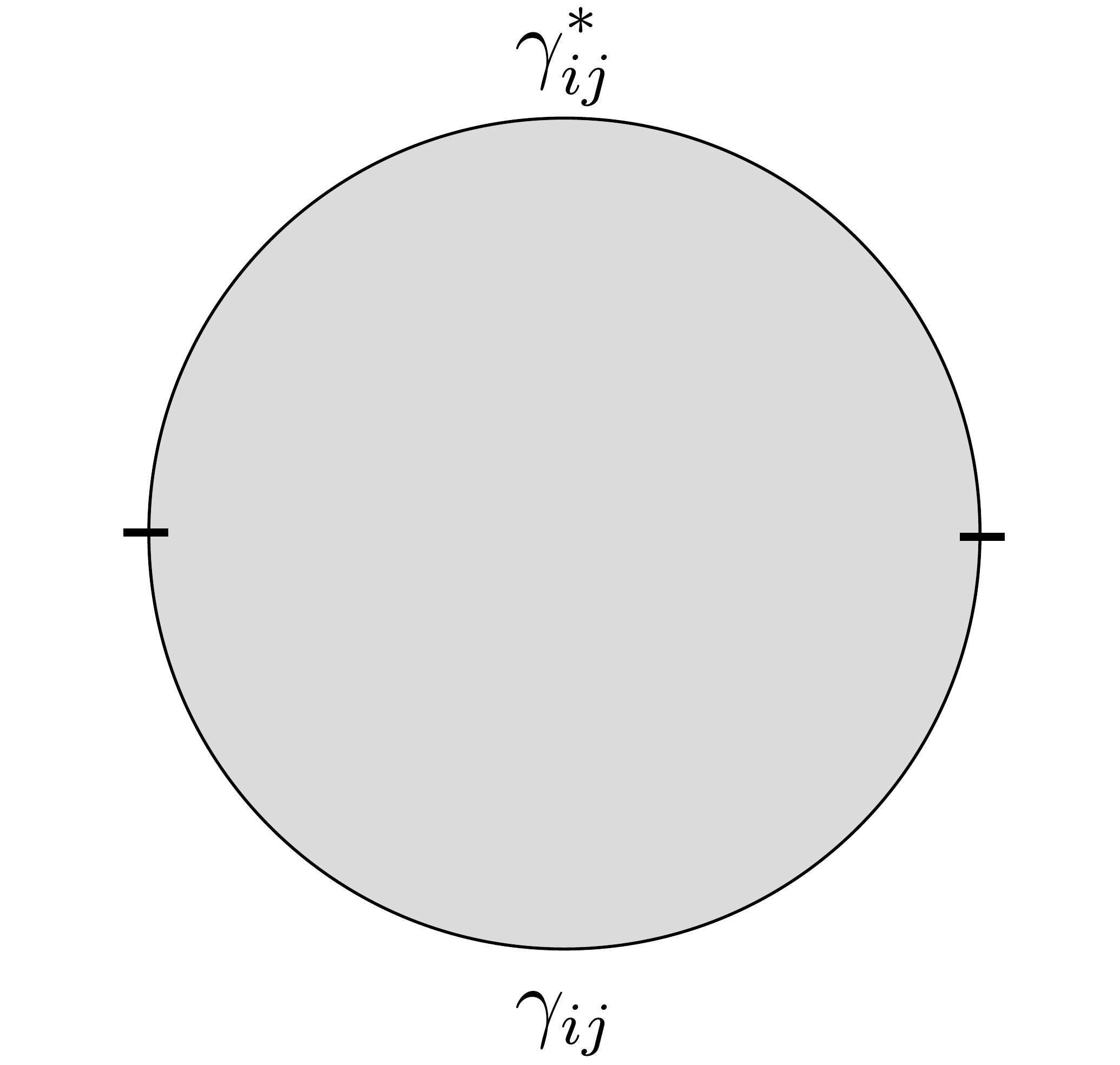}
\includegraphics[width=0.3\textwidth]{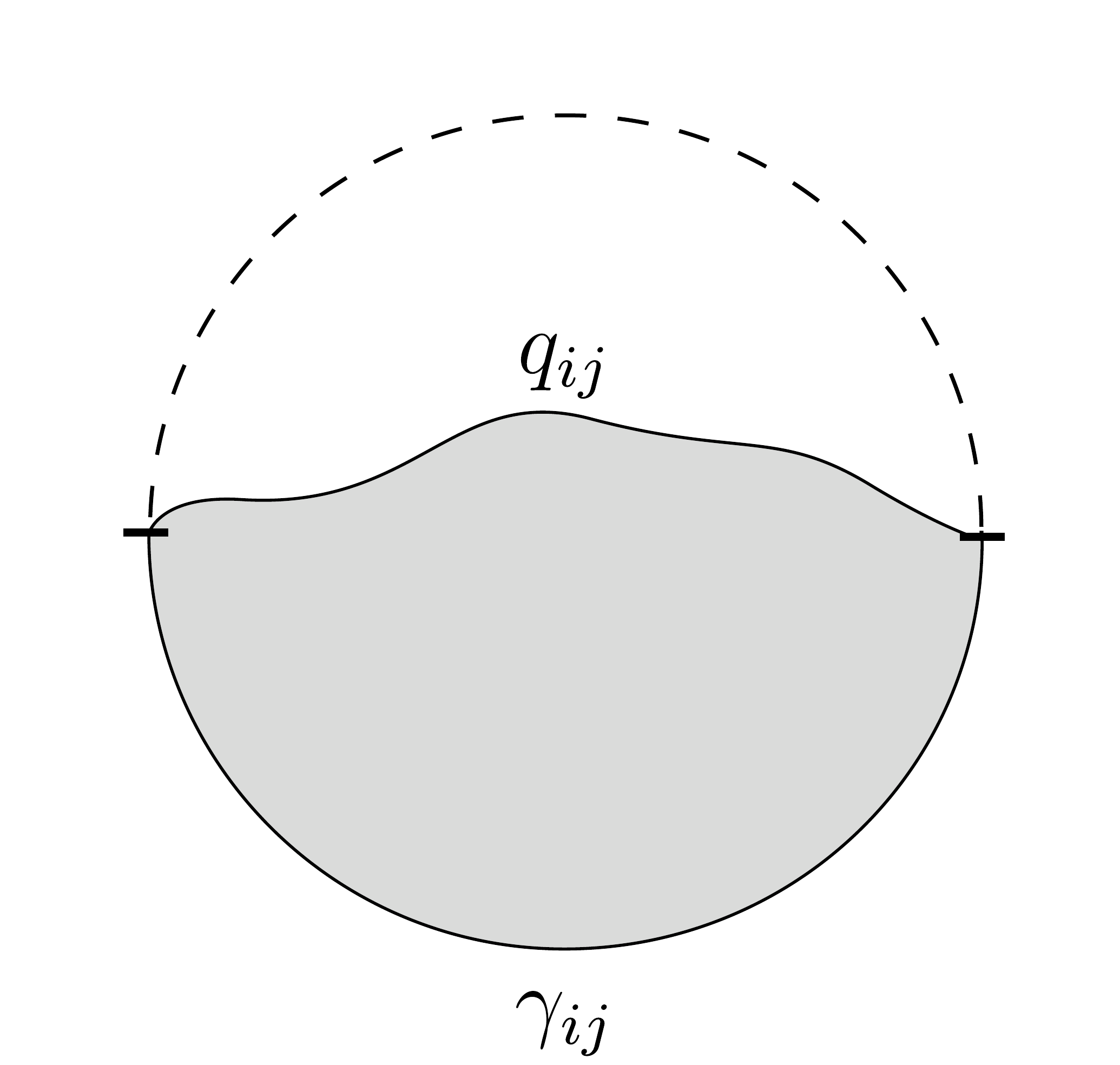}
\includegraphics[width=0.3\textwidth]{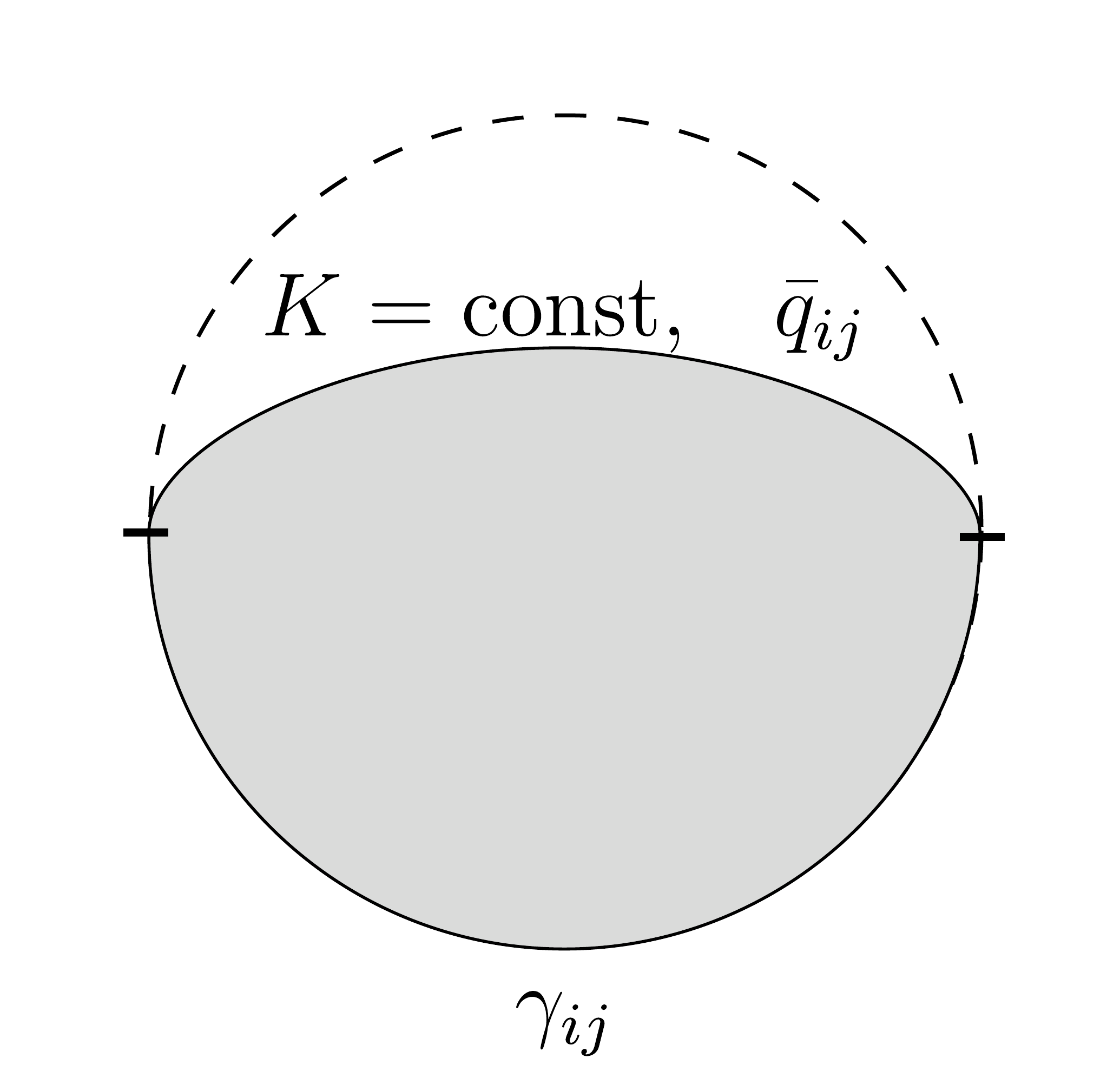}
\caption{Three different Dirichlet problems.}
\label{fig:overlaps}
\end{center}
\end{figure}

First, recall that the partition function of the CFT with position dependent sources turned on, can be regarded as an overlap of two states, see left of Fig. \ref{fig:overlaps}. We will restrict attention to the case where we only source the stress tensor with a background metric. In this interpretation, the metric on the lower hemisphere prepares a state and the metric on the upper hemisphere prepares the conjugate of a state. To obtain arbitrary bulk initial data, one must complexify this metric \cite{Marolf:2017kvq,Belin:2018fxe,Belin:2018bpg}. Such overlaps are therefore related to the (renormalized) partition function 
\beq
\label{eq:partitionoverlap}
\langle \gamma_1|\gamma_2\rangle = Z^{\rm CFT}_{\rm ren}[\Theta({\rm up})\gamma_1^* + \Theta({\rm down})\gamma_2],
\eeq
where the $*$ includes an Euclidean time reflection along with the conjugation of the source. Here we denote by $|\gamma\rangle$ an unnormalized Euclidean path integral state and $\gamma_{1,2}$ are metrics on the hemisphere. In order to define finite energy states in the CFT, we need to require certain falloff conditions near the $t=0$ slice where the overlap is glued. Since (up to conformal anomaly) only the conformal class matters, we may enforce this by requiring $\gamma_{1,2}$ to be asymptotically $\mathbb{H}_{d}$. The vacuum corresponds to the standard hyperbolic metric on $\mathbb{H}_{d}$ and gives rise to Euclidean AdS$_{d+1}$ in the bulk in the slicing
\beq
\label{eq:AdSCMC}
ds^2=d\tau^2+\cosh^2\tau d\Sigma_{\mathbb{H}_{d}}^2,
\eeq
where $\tau\rightarrow \pm \infty$ are the northern/southern hemispheres of the boundary. The asymptotic boundary of the $\mathbb{H}_{d}$ slices is the $t=0$ slice in the CFT where we glue the overlap.

Let us denote the Hartle-Hawking wavefunction associated to the state $|\gamma\rangle$ by $\Phi_{\gamma}[q]$. This wavefunction is computed in the bulk by performing the Euclidean path integral between half of the conformal boundary with metric $\gamma_{ij}$ and another co-dimension one surface with metric $q_{ij}$, see the middle figure on Fig. \ref{fig:overlaps}. These are both Dirichlet boundary conditions. In addition, we should include holographic counter terms for the boundary condition imposed at conformal infinity in order to obtain a finite result. This prescription makes it clear that the ``radial" wavefunction encodes the wave functions of path integral states in the CFT via
\beq
\label{eq:HHwavefunction}
\Phi_{\gamma}[q_{ij}] \propto \lim_{\lambda \rightarrow 0}  e^{-S_{c.t}[\gamma/\lambda^{2/d}]}\Psi[\Theta({\rm up})q_{ij}+\Theta({\rm down})\gamma_{ij}/\lambda^{2/d}],
\eeq
where the counter term action $S_{c.t}$ is integrated only over the lower hemisphere.\footnote{Here and bellow we restrict formulas to odd boundary dimensions where $S_{c.t.}$ is a function of $\gamma/\lambda^{2/d}$. In even dimensions, there is a $\log \lambda$ term encoding the conformal anomaly. As usual, for us, $S_{c.t.}$ contains only the counter terms involving derivatives.} We may also interpret this formula as having a single $\lambda(x)=\Theta({\rm up})+\Theta({\rm down})\lambda$ that is position dependent and therefore we can also express the r.h.s. with the bare partition function \eqref{eq:frompartfunctowavefunc}
\beq
\Phi_{\gamma}[q_{ij}] \propto \lim_{\lambda \rightarrow 0}  e^{-S_{c.t}[\gamma/\lambda^{2/d}]}Z^{(\Theta({\rm up})+\Theta({\rm down})\lambda)}[\Theta({\rm up})q_{ij}+\Theta({\rm down})\gamma_{ij}].
\eeq

It is also possible to formally define bulk ``metric eigenstates" as linear combinations of the path integral states $|\gamma\rangle$. Let us denote such a metric eigenstate $|q\rangle \rangle$ in order to distinguish it from a path integral state. We define these states via
\beq
\Phi_{\gamma}[q_{ij}]=\langle \langle q_{ij}|\gamma\rangle.
\eeq
Because of \eqref{eq:HHwavefunction} and \eqref{eq:partitionoverlap}, this wave function can be though of as a hybrid bare-renormalized partition function that has been flowed with the $T^2$ flow equation \eqref{eq:T2flow1} only on the northern hemisphere, hence it is related to a CFT partition function by the application of a diffusion kernel
\bea
\label{eq:flowHH}
\langle \langle q_{ij}|\gamma\rangle &= \int \mathcal{D}\gamma' {\tilde L}[q,\gamma'] Z^{\rm CFT}_{\rm ren}[\Theta({\rm up})\gamma' + \Theta({\rm down})\gamma] \\
&\equiv \int \mathcal{D}\gamma' {\tilde L}[q,\gamma'] \langle \gamma'|\gamma\rangle.
\eea
Here, $\tilde L$ is related to the diffusion kernel solving \eqref{eq:T2flow1} and described in sections \ref{sec:nopotkernel},\ref{sec:addingRicci}, by a dressing with counter terms on the right
\beq
{\tilde L}[q,\gamma'] = \lim_{\lambda \rightarrow 0}L[q,\gamma'/\lambda^{2/d}]e^{S_{c.t.}[\gamma'/\lambda^{2/d}]}
\eeq
so that it acts on renormalized partition functions, transforming them into bare ones. Formally, \eqref{eq:flowHH} is equivalent with writing
\beq
\label{eq:basistrf}
|q_{ij}\rangle \rangle = \int \mathcal{D}\gamma' {\tilde L}[q,\gamma'] |\gamma'\rangle,
\eeq
which is the announced linear relation between path integral states and metric eigenstates. There is a very important point to stress regarding this formula. The path integral calculating the overlap \eqref{eq:flowHH} is dominated by a classical saddle solving bulk equations of motion in the $G_N\rightarrow 0$ limit. This is certainly \textit{not} true for equation \eqref{eq:basistrf}. Saddles may dominate in calculating the overlap of  \eqref{eq:basistrf} with certain states, but the saddle will highly depend on which states we are calculating the overlap with.

\subsection{York time and Lorentzian wave functions}

The wave function \eqref{eq:HHwavefunction} is still an intrinsically Euclidean object and one may wonder how it is related to Lorentzian wave functions. In particular, since the slice on which $\Phi_\gamma$ is defined is determined by the scale argument, we should really think about it as being analogous to a time dependent Schr\"odinger wave function, that requires Wick rotation.\footnote{For example, under $\lambda \rightarrow i\lambda$, the $T^2$ flow equation \eqref{eq:T2flow1} becomes a Schr\"odinger type equation. This continuation is not relevant for AdS/CFT but might be relevant for wave functions in de Sitter.} A natural approach that we will use here is based on York time \cite{York:1972sj}. York time is defined via slicing a geometry into constant mean curvature (CMC) slices, parametrized by their constant trace of extrinsic curvature $K\equiv K_{ij}q^{ij}=\text{const}$. This is expected to be a well defined slicing of the causal diamond anchored at a fixed boundary time slice in the case of Lorentzian asymptotically AdS geometries, with $K$ being a well-defined notion of time in the diamond \cite{witten2017}. For empty AdS, this slicing is given by Wick rotating \eqref{eq:AdSCMC}. Therefore, it seems natural to Wick rotate wave functions along the extremal slice $K=0$ by sending $K\rightarrow iK$. Of course, the wave function \eqref{eq:HHwavefunction} depends on the metric instead of the extrinsic curvature. It is better therefore to switch representation and consider the Laplace transform with respect to the scale \cite{Hartle:1983ai}
\beq
\label{eq:Yorkwavefunc}
\Phi_{\gamma}[K,{\bar q}_{ij}] \sim \int_{0}^\infty \mathcal{D}[\sqrt{q}]e^{-\frac{1}{d}\int \sqrt{q}\theta}\Phi_\gamma[q_{ij}]
\eeq
where ${\bar q}_{ij}=|q|^{1/d}{ q}_{ij}$ is the conformal metric and $\theta=2(1-d)[K+d/\ell]$ is the trace of the bare stress tensor.\footnote{We mean here the stress tensor coming from the bare partition function. As explained, this still contains the leading holographic counter term, so the formula is $T^{ij}_{\rm bare}=2(K^{ij}-K q^{ij})-2\frac{d-1}{\ell} q^{ij}$, where we have set $16\pi G_N=1$ as before. The role of the shift in the exponent compared to $K$ is to remove the leading holographic counter term from $\Phi_\gamma[q_{ij}]$, since as we have defined it, $\Phi_\gamma[q_{ij}]$ solves the canonically transformed WDW equation \eqref{eq:WDW0}.} Notice that the inverse Laplace transform to \eqref{eq:Yorkwavefunc} is naturally over Lorentzian values of $K$. In the semiclassical limit, $\Phi_{\gamma}[K,{\bar q}_{ij}]$ is calculated by an on-shell action with Dirichlet condition on $K$ instead of $\sqrt{q}$, which is arguably a better defined problem in Euclidean signature \cite{Witten:2018lgb}. Note that this object is related to the Laplace transform of the $T^2$ flowed partition function with respect to the inverse of the flow parameter $u=1/\lambda$ restricted to the upper hemisphere, since
\bea
\label{eq:laplacebare}
\lim_{u_{\rm down}\rightarrow \infty}\int \mathcal{D} & u_{\rm up} e^{-\int \sqrt{\gamma_{\rm up}} u_{\rm up} \theta} Z^{(1/u)}[\gamma] \\ & =
\lim_{u_{\rm down}\rightarrow \infty}\int \mathcal{D}u_{\rm up} e^{-\int \sqrt{\gamma_{\rm up}} u_{\rm up} \theta} \Psi[\Theta({\rm up})\gamma_{\rm up} u_{\rm up}^{2/d} + \Theta({\rm down})\gamma_{\rm down} u_{\rm down}^{2/d}]\\
&\propto \int \mathcal{D}\left( \frac{\sqrt{q}}{d\sqrt{\gamma_{\rm up}}}\right) e^{-\frac{1}{d}\int \sqrt{q}\theta} \Phi_{\gamma_{\rm down} }[q_{ij}],\\
&\propto \Phi_{\gamma_{\rm down}}[K,{\bar q}_{ij}] ,
\eea
where we used \eqref{eq:frompartfunctowavefunc}, \eqref{eq:HHwavefunction} and changed integration variable. Therefore this wave function will be related to solutions of the Laplace transformed flow equation \eqref{eq:T2flow1}. In the spirit of the CMC slicing, we imagine evaluating $\Phi_{\gamma}[K,{\bar q}_{ij}]$ for constant $K$ and thinking about it as a Schr\"odinger wave function with time $K$.

\subsection{Volume of the maximal slice}

The Hamiltonian conjugate to York time which moves between CMC slices is the volume \cite{York:1972sj}. In the semiclassical limit, the wave function \eqref{eq:Yorkwavefunc} will be given by an on-shell action which is a boundary term appropriate for fixing $K$
\beq
\Phi_{\gamma}[K,{\bar q}_{ij}] \sim e^{\frac{2}{d} \int \sqrt{q}K},
\eeq
where the volume density $\sqrt{q}$ is some functional of $K$, ${\bar q}_{ij}$ and $\gamma_{ij}$ determined by solving the equation of motion. It follows that
\beq
\label{eq:volumeformula}
\partial_K\log\Phi_{\gamma}[K,{\bar q}_{ij}]|_{K=0}=\frac{2}{d}V_{\rm ex},
\eeq
where $V_{\rm ex}$ is the volume of the extremal slice. Since $\Phi_{\gamma}[K,{\bar q}_{ij}]$ may be interpreted as a certain Laplace transform of a $T^2$ flowed bare partition function via \eqref{eq:laplacebare}, one may formally interpret this as an expression solely in terms of boundary data:
\beq
\label{eq:volumeformula2}
\lim_{u_{\rm down}\rightarrow \infty}\partial_\theta {\log\left[ \int \mathcal{D}  u_{\rm up} e^{-\int \sqrt{\gamma_{\rm up}} u_{\rm up} \theta} Z^{(1/u)}[\gamma] \right]\vline}_{\; \theta=2(1-d)d/\ell} = \frac{1}{d(1-d)}V_{\rm ex}.
\eeq
Of course, in practice we do not find this particularly useful, since the only way we know how to evaluate such an object is via solving the gravity equations of motion. It also requires some definition of a local $T^2$ deformation that is only available at large $N$. For finite $N$, we only know examples of such well defined deformations for 2d CFTs, and even in that case not for arbitrary background metric, see however \cite{Jiang:2019tcq,Mazenc:2019cfg,Brennan:2020dkw} for progress in this direction. 

Nevertheless, it would be interesting to try to interpret \eqref{eq:volumeformula2} in terms of the complexity=volume proposal \cite{Susskind:2014rva,Stanford:2014jda}, perhaps by trying to interpret the Laplace transformed $T^2$ flow equation as a solution to some optimization problem. It would also be interesting to see if there is a connection to the tensor network interpretation of $T\bar T$ put forward in \cite{Kruthoff:2020hsi} (see also \cite{Geng:2019yxo}). We leave these questions for future work.

\section{Discussion}
\label{sec:discussion}

In this section, we briefly discuss some open questions that were raised along the way.

\subsection{Uniqueness of $T^2$ flow}

Unlike in $d=2$ on a flat geometry, it is unclear whether the $T^2$ deformation makes sense at the full-blown quantum level. The definition of the $T^2$ operator which induces the flow relied crucially on the large $N$ limit and can be defined in the $1/N$ expansion. From an abstract CFT point of view, the key feature used in \cite{Hartman:2018tkw} was large $N$-factorization, namely that
\be
\braket{:T^2:}_{\ket{\psi}}=\braket{T}_{\ket{\psi}}\braket{T}_{\ket{\psi}}+\mathcal{O}\left(\frac{1}{N}\right) \,.
\ee
It is important to note that this property will be obeyed in any large $N$ CFT with a 't Hooft expansion, irrespective of whether the theory is actually dual to Einstein gravity (i.e. a 2-derivative theory) or not. Said differently, the factorization property of the $T^2$ operator will hold even if the CFT does not have a large gap \cite{Heemskerk:2009pn,Afkhami-Jeddi:2016ntf,Belin:2019mnx,Kologlu:2019bco}, in which case higher derivative corrections become important.

This immediately suggests that the flow into the bulk should not be unique in any dimension $d>2$, and the details of the deformation operator should encode the nature of the gravitational theory. To the best of our knowledge, the flow equation corresponding to a higher derivative theory has not been worked out and it would be interesting to do so. A connection to this question has already appeared in our derivation of the kernel: To obtain a $d+1$-dimensional diff-invariant action, we added a Ricci-potential that combined with the extrinsic curvature squared into the $d+1$-dimensional Ricci scalar, i.e. the Einstein Hilbert action. One can ask whether this was the unique way to produce a scalar? In particular, it is clear that there are other potentials preserving the symmetry \rref{eq:potentialscaling}. We believe that without adding higher order terms in the extrinsic curvature, it is (up to just changing the cosmological constant), and that adding higher order extrinsic curvatures would precisely amount to higher derivative theories. This suggests that higher-derivative flows would correspond to having higher powers of the stress-tensor in the deformation, which sounds intriguing. We hope to return to this question in the future.

\subsection{A fake bulk?}

A related puzzle arises from noticing that we have nowhere used any assumption about the boundary CFT besides that the $:T^2:$ operator is well defined, i.e. large $N$.
So what happens if we take a large $N$ theory, and flow with the wrong operator? For example, we could deform weakly-coupled $\mathcal{N}=4$ SYM with the Einstein gravity $T^2$ operator. Even at strong coupling, if there are single trace operators besides the stress tensor, the appropriate double traces must be added to the deforming operator to generate the right bulk \cite{Kraus:2018xrn}. In the absence of these, one will just generate the pure gravitational path integral. So a natural question is how to tell if we are using the right double trace deformation, or that the bulk we are flowing into is not ``fake".\footnote{Another way of changing the deforming operator is by adding an extra $\frac{1}{\lambda^2}\sqrt{|\gamma|}$ term to the potential in \eqref{eq:flowwithgeneralpot}, which is also consistent with the scaling property \eqref{eq:potentialscaling}. This will change the cosmological constant generated in the kernel, in particular, one can use this to flip its sign. For the $d=2$ case this was explored in \cite{Gorbenko:2018oov}. One may worry that the bulk generated this way is ``fake" in the same sense as the AdS bulk is fake when we flow from a weakly coupled theory.} We have no definite answer to this question, but we offer some speculation. 

A practical condition is to require that the generated path integral is consistent with GKPW \cite{Gubser:1998bc,Witten:1998qj}. If we put the CFT on a manifold with spherical topology, the diffusion kernel will be given by a gravitational path integral between two spheres. In order to recover that the CFT partition function is a gravity path integral with only one spherical boundary, we must require that at the limit of zero scale $\sqrt{q}\rightarrow 0$ the radial wave function is trivial $\Psi\rightarrow 1$. This is the limit of large flow parameter $\lambda \rightarrow \infty$ for the partition function. In general, we could land on any functional of the conformal metric ${\bar q}_{ij}$ in this limit. Where we land depends on the initial condition at $\sqrt{q}\rightarrow \infty$, i.e. the CFT partition function that we deform. Having $\lim_{\lambda \rightarrow \infty}\Psi[\gamma_{ij}/\lambda^{2/d}]\propto 1$ seems to be a nontrivial condition on this initial data. It would be interesting to explore this (and other possible answers to this question) further.

\subsection{Connection to the Freidel kernel}

As mentioned in the introduction, the solution of the $T\bar T$ flow in $d=2$ in terms of a diffusion kernel is known. This kernel was first obtained by Freidel in \cite{Freidel:2008sh} in the context of the WDW equation and was later utilized for the $T\bar T$ flow in \cite{McGough:2016lol,Mazenc:2019cfg}. However, the Freidel kernel is a simple exponential of a local functional of the veilbeins of the initial and final 2d metrics, as opposed to being given by a path integral in 3d gravity. Note that in $d=2$ our deforming operator \eqref{eq:Hartmanflow} is just $X=T\bar{T}$ without any potential terms, therefore we are solving the same flow equation as \cite{McGough:2016lol,Mazenc:2019cfg}, which implies that our kernel \eqref{eq:pathintkernel} (with the log counter term included) must agree with the Freidel kernel. It is known that the path integral of 3d gravity in AdS, formulated as a Chern-Simons theory \cite{Witten:1988hc}, localizes on the boundary, essentially because the temporal component of the Chern-Simons field acts as a Lagrange multiplier \cite{Elitzur:1989nr,Coussaert:1995zp}. Using this, it might be possible to explicitly confirm that our result reduces to the simple form of the Freidel kernel in $d=2$. It would be interesting to do so.

\section*{Acknowledgements}

We thank V. Balasubramanian, P. Caputa, Y. Jiang, J. Kruthoff, O. Parrikar, E. Shaghoulian and A. Wall for fruitful discussions. A.B. and G.S. would like to thank A.L. and his collaborators for finishing their other projects in time, so that this can be A.L.'s last hep-th paper. The work of A.B. is supported in part by the NWO VENI grant 680-47-464 / 4114.

\appendix

\section{Scaling Ward identity for the renormalized partition function}
\label{app:anomaly}

Here we argue that the anomaly free Ward identity \eqref{eq:scaling} for the bare partition function is consistent with the usual anomalous Ward identity for the renormalized stress tensor in even dimensions. In other words, when the $T^2$ coupling is upgraded to a local background field $\lambda(x)$ on which Weyl rescalings act, the Weyl anomaly may be removed by a local counterterm.

In terms of the renormalized partition function, using \eqref{eq:ct1} and \eqref{eq:bare-renorm}, the Ward identity \eqref{eq:scaling} reads as 
\beq
\label{eq:renormward}
Z_{\rm ren}^{(\alpha^{-d}\lambda)}[\alpha^{-2}\gamma]=e^{\mathcal{A}[\alpha^{-d}\lambda,\alpha^{-2}\gamma]-\mathcal{A}[\lambda,\gamma]}Z_{\rm ren}^{(\lambda)}[\gamma].
\eeq
where
\beq
\mathcal{A}[\lambda,\gamma]=\frac{1}{d}\int \log \lambda\sqrt{\gamma}A_d[ \gamma].
\eeq
The usual form of the identity is obtained by taking an $\alpha(x)$ variation of the log of \eqref{eq:renormward}, setting $\alpha=1$ and using \eqref{eq:Xdef}:
\bea
\label{eq:localWard}
(\Delta_X-d)\lambda \langle X\rangle+\langle T^i_i\rangle &=\frac{1}{\sqrt{\gamma}}\left[d \lambda \frac{\delta \mathcal{A}}{\delta \lambda}+2 \gamma_{ij}\frac{\delta \mathcal{A}}{\delta \gamma_{ij}} \right]
\\&= A_d[\gamma]+(T_{\mathcal{A}})^i_i ,
\eea
where $\Delta_X=2d$ and $(T_{\mathcal{A}})^{ij}=\frac{2}{\sqrt{\gamma}}\frac{\delta \mathcal{A}}{\delta \gamma_{ij}}$ is a stress tensor arising from the action $\mathcal{A}$. When $\lambda$ is constant, $(T_{\mathcal{A}})_{ij}=d^{-1}(\log \lambda) T^a_{ij}$, where $T^a_{ij}$ is the anomaly stress tensor defined in \cite{deHaro:2000vlm}, in particular it is proportional to the coefficient of the $\log$ term in the FG expansion of the bulk metric, which is denoted in \cite{deHaro:2000vlm} by $h_{(d)ij}$. It is shown in \cite{deHaro:2000vlm} that this term is traceless when contracted with the coefficient of the leading term, i.e. the CFT background metric, so we have $\gamma^{ij}h_{(d)ij}=0$. This implies that for constant $\lambda$, $(T_{\mathcal{A}})^i_i=0$ in \eqref{eq:localWard}, and the anomaly is given by the usual holographic Weyl anomaly \cite{Henningson:1998gx}. In the case of non-constant $\lambda$, $(T_{\mathcal{A}})_{ij}$ contains terms proportional to $\partial_k \log \lambda$ in addition to $d^{-1}(\log \lambda) T^a_{ij}$. These terms are coming from integrating by parts derivatives of $\delta \gamma_{ij}$. The trace of these terms gives a possibly nonzero contribution to $(T_{\mathcal{A}})^i_i$ which must be included in \eqref{eq:localWard} away from the conformal point $\lambda=0$.

\bibliographystyle{utphys}
\bibliography{volume}

\end{document}